# Inapproximability of Nash Equilibrium

Aviad Rubinstein*

September 9, 2016

**Abstract**

We prove that finding an $\epsilon$-approximate Nash equilibrium is PPAD-complete for constant $\epsilon$ and a particularly simple class of games: polymatrix, degree 3 graphical games, in which each player has only two actions.

As corollaries, we also prove similar inapproximability results for Bayesian Nash equilibrium in a two-player incomplete information game with a constant number of actions, for relative $\epsilon$-Well Supported Nash Equilibrium in a two-player game, for market equilibrium in a non-monotone market, for the generalized circuit problem defined by Chen et al [CDT09], and for approximate competitive equilibrium from equal incomes with indivisible goods.

## 1 Introduction

Nash equilibrium is the central concept in Game Theory. Much of its importance and attractiveness comes from its *universality*: by Nash's Theorem [Nas51], every finite game has at least one. The result that finding a Nash equilibrium is PPAD-complete, and therefore intractable [DGP09, CDT09] casts this universality in doubt, since it suggests that there are games whose Nash equilibria, though existent, are in any practical sense inaccessible.

*UC Berkeley. I am grateful to Christos Papadimitriou for inspiring discussions, comments, and advice. I also thank Yang Cai for suggesting the connection to Bayesian Nash equilibrium in two-player games. Additional thanks go to Constantinos Daskalakis, Karthik C.S., and anonymous reviewers for helpful comments. This research was supported by NSF grants CCF0964033 and CCF1408635, and by Templeton Foundation grant 3966. This work was done in part at the Simons Institute for the Theory of Computing.

Can approximation repair this problem? Chen et al [CDT09] proved that it is also hard to find an $\epsilon$-approximate Nash equilibrium for any $\epsilon$ that is polynomially small - even for two-player games. The only remaining hope is a PTAS, i.e. an approximation scheme for any constant $\epsilon > 0$. *Whether there is a PTAS for the Nash equilibrium problem is the most important remaining open question in equilibrium computation.*

For a constant number of players, we are unlikely to prove PPAD-hardness since a quasi-polynomial time approximation algorithm exists [LMM03][1]. We therefore shift our attention to games with a large number of players. For such games, there is a question of representation: the normal form representation is exponential in the number of the players. Instead, we consider two natural and well-studied concise representations:

**Polymatrix games** In a polymatrix game, each pair of players simultaneously plays a separate two-player game. Every player has to play the same strategy in every two-player subgame, and her utility is the sum of her subgame utilities. The game is given in the form of the payoff matrix for each two-player game.

**Graphical games** In a graphical game [Kea07], the utility of each player depends only on the action chosen by a few other players. This game now naturally induces a directed graph: we say that $(i, j) \in E$ if the utility of player $j$ depends on the strategy chosen by player $i$. When the maximal incoming degree is bounded, the game has a representation polynomial in the number of players and strategies.

### Our results

We prove that even for games that are both polymatrix and graphical (for particularly simple graphs) finding an $\epsilon$-approximate Nash equilibrium is intractable.

**Theorem 1.** *There exists a constant $\epsilon > 0$, such that given a degree 3, bipartite, polymatrix game where each player has two actions, finding an $\epsilon$-approximate Nash equilibrium is* PPAD*-complete.*

In previous works [CDT09, DGP09], hardness of Nash equilibrium in two-player games was obtained from hardness of a bipartite polymatrix game. Roughly speaking, the reduction lets each of the two players choose the strategies for the vertices on one side of the bipartite graphical game. This reduction incurs a polynomial blowup in the error - and indeed, as discussed earlier, we do not expect to obtain PPAD-hardness for $\epsilon$-approximate Nash equilibrium in two-player games. Nevertheless, in Section 8 we show that a similar technique yields an interesting corollary for $\epsilon$-approximate Bayesian Nash equilibrium in two-player games with incomplete information.

[1] In followup work [Rub16] we prove that assuming the "Exponential Time Hypothesis (ETH) for PPAD", quasi-polynomial running time of [LMM03]'s algorithm is indeed necessary.

**Corollary 1** ($\epsilon$-Bayesian Nash equilibrium)**.** *There exists a constant $\epsilon > 0$, such that given a two-player game with incomplete information where each player has a constant number of actions, finding an $\epsilon$-approximate Bayesian Nash equilibrium is* PPAD*-complete.*

In two-player complete information games, Daskalakis [Das13] circumvents Lipton et al's quasi-polynomial time algorithm by studying a notion of relative (sometimes also called *multiplicative* [HdRS08], as opposed to the more standard additive) $\epsilon$-Well Supported Nash Equilibrium (relative $\epsilon$-WSNE). Daskalakis proves that in two player games with payoffs in $[-1, 1]$, finding a relative $\epsilon$-WSNE is PPAD-complete. One caveat of this result is that the gain from deviation is large compared to the expected payoff because the latter is small due to cancellation of positive and negative payoffs. Namely, the gain from deviation may be very small compared to the expected magnitude of the payoff. Here we answer an open question from [Das13] by proving that finding a relative $\epsilon$-WSNE continues to be PPAD-complete even when all the payoffs are positive.

**Corollary 2** (Relative $\epsilon$-WSNE)**.** *There exists a constant $\epsilon > 0$ such that finding a relative $\epsilon$-Well Supported Nash Equilibrium in a bimatrix game with positive payoffs is PPAD-complete.*

The computation of Nash equilibrium is tightly related to computation of equilibrium in markets. In particular, Chen, Paparas, and Yannakakis [CPY13] use a reduction from polymatrix games to prove PPAD-hardness for the computation of equilibrium for every family of utility functions from a very general class, which they call *non-monotone families of utility functions.* In Section 10 we prove the following hardness of approximation for market equilibrium.

**Corollary 3** (Non-monotone markets)**.** *Let $\mathcal{U}$ be any non-monotone family of utility functions. There exists a constant $\epsilon_{\mathcal{U}} > 0$ such that given a market $M$* *where the utility of each trader is either linear or taken from $\mathcal{U}$, finding an $\epsilon_{\mathcal{U}}$-tight approximate market equilibrium is* PPAD*-hard.*

Although our inapproximability factor is stronger than that showed by Chen et al, the results are incomparable as ours only holds for the stronger notion of "tight" approximate equilibrium, by which we mean the more standard definition which bounds the two-sided error of the market equilibrium. Chen et al, in contrast, prove that even if we allow arbitrary excess supply, finding a $(1/n)$-approximate equilibrium is PPAD-hard. Furthermore, for the interesting case of CES utilities with parameter $\rho < 0$, they show that there exist markets where every $(1/2)$-tight equilibrium requires prices that are doubly-exponentially large (and thus require an exponential-size representation). Indeed, for a general non-monotone family $\mathcal{U}$, the problem of computing a (tight or not) approximate equilibrium may not belong to PPAD. Nevertheless, the important family of additively separable, concave piecewise-linear utilities is known to satisfy the non-monotone condition [CPY13], and yet the computation of (exact) market equilibrium is in PPAD [VY11]. Therefore,

**Corollary 4** (SPLC markets)**.** *There exists a constant $\epsilon > 0$, such that finding an $\epsilon$-tight approximate market equilibrium with additively separable, concave piecewise-linear utilities is* PPAD*-complete.*

En route to proving our main result, we also prove hardness of approximation for the generalized circuit problem. Generalized circuits are similar to standard algebraic circuits, the main difference being that generalized circuits contain cycles, which allow them to verify fixed points of continuous functions. A generalized circuit induces a constraint satisfaction problem, $\epsilon$-GCIRCUIT [CDT09]: find an assignment for the values on the lines of the circuit, that simultaneously $\epsilon$-approximately satisfies all the constraints imposed by the gates (see Section 3.2 for a formal definition). $\epsilon$-GCIRCUIT was implicitly proven PPAD-complete for exponentially small $\epsilon$ by Daskalakis et al [DGP09], and explicitly for polynomially small $\epsilon$ by Chen et al [CDT09]. Here we prove that it continues to be PPAD-complete for some constant $\epsilon$.

**Theorem 2** (Generalized circuit)**.** *There exists a constant $\epsilon > 0$ such that $\epsilon$-*GCIRCUIT *with fan-out* 2 *is* PPAD*-complete.*

The $\epsilon$-GCIRCUIT problem has already proven useful in several works in recent years (e.g [CDT09, Das13, CPY13, OPR14]). We believe that Theorem 2 will lead to stronger hardness results in many applications in algorithmic game theory and economics. For example, competitive equilibrium with equal incomes (CEEI) is a well-known fair allocation mechanism [Fol67, Var74, TV85]; however, for indivisible resources a CEEI may not exist. It was shown by Budish [Bud11] that in the case of indivisible resources there is always an allocation, called A-CEEI, that is approximately fair, approximately truthful, and approximately efficient, for some favorable approximation parameters. This approximation is used in practice to assign business school students to classes [Oth14].

The approximation of CEEI is characterized by two parameters: $\alpha$, which is a measure of the market clearing error, and $\beta$, which quantifies the inequality in initial endowments. Budish's proof guarantees the existence of an $(\alpha^*, \beta)$-CEEI for any $\beta > 0$ (and some parameter $\alpha^*$ that depends on the input). Othman et al [OPR14] reduced $\Theta(\beta \log(1/\beta))$-GCIRCUIT with fan-out 2 to the problem of finding the guaranteed approximation, $(\alpha^*, \beta)$-CEEI. Theorem 2 now gives the following corollary:

**Corollary 5** (A-CEEI). *There exists a constant $\beta > 0$ such that finding an $(\alpha^*, \beta)$-CEEI is* PPAD*-complete.*

## 1.1 Related works

This work extends [DGP09, CDT09, CPY13, OPR14] where similar hardness results were established for $1/\mathsf{poly}(n)$-approximation. It also extends our own [Rub14], where we proved PPAD-hardness of $\epsilon$-Well Supported Nash equilibrium ($\epsilon$-WSNE), for constant $\epsilon$ and a constant number of actions per player, over a larger class of games called *succinct games*. Succinct games can have much more complex utilities than polymatrix or graphical games of bounded degree: any utility function that can be computed in polynomial time is allowed.

We improve over the results of [Rub14] in three ways: (1) simplicity: our new hardness result holds for a much simpler and more natural definition of games; (2) $\epsilon$-ANE vs $\epsilon$-WSNE: thanks to the transformation to bounded degree graphical games our result extends to the weaker concept (thus stronger hardness) of $\epsilon$-*approximate* Nash equilibrium (see Section 3.1 for precise definitions); and (3) completeness: finding an approximate Nash equilibrium in a bounded-degree graphical game is not only PPAD-hard - it also belongs to PPAD [DGP09]. For succinct games, in contrast, $\epsilon$-WSNE is unlikely to belong to PPAD since it is also known to be BPP-hard [SV12].

On the technical side, our current construction of hard instances of Brouwer fixed points is identical to [Rub14]. The main technical contribution in this paper is the adaptation of the equiangle sampling gadget of Chen et al [CDT09] to this particular Brouwer function.

**Query complexity** There are several interesting results on the query complexity of approximate Nash equilibria, where the algorithm is assumed to have black-box access to the exponential-size payoff function.

Hart and Nisan [HN13] prove that any deterministic algorithm needs to query at least an exponential number of queries to compute any $\epsilon$-Well Supported Nash Equilibrium - and even for any $\epsilon$-correlated equilibrium. For $\epsilon$-correlated equilibrium, on the other hand, Hart and Nisan show a randomized algorithm that uses a number of queries polynomial in $n$ and $\epsilon^{-1}$.

Babichenko [Bab14] showed that any randomized algorithm requires an exponential number of queries to find an $\epsilon$-Well Supported Nash Equilibrium. Our proof is inspired by Babichenko's work and builds on some of his techniques. Babichenko's query complexity lower bound was eventually extended to $\epsilon$-Approximate Nash Equilibrium [CCT15, Rub16] and later also to communication complexity [BR16].

Goldberg and Roth [GR14] characterize the query complexity of approximate coarse correlated equilibrium in games with many players. More important for our purpose is their polynomial upper bound on the query complexity of $\epsilon$-WSNE for any family of games that have *any* concise representation. This result is to be contrasted with (a) Babichenko's query complexity lower bound, which uses a larger family of games, and (b) [Rub14] which applies exactly to this setting and gives a lower bound on the *computational complexity*.

A much older yet very interesting and closely related result is that of Hirsch, Papadimitriou, and Vavasis [HPV89]. Hirsch et al show that any deterministic algorithm for computing a Brouwer fixed point in the oracle model must make an exponential -in the dimension $n$ and the approximation $\epsilon$- number of queries for values of the function. The techniques in [HPV89] have proven particularly useful both in [Bab14, Rub14] and here.

**Approximation algorithms** In any polymatrix game and for any constant $\delta > 0$, Deligkas et al [DFSS14] give a polynomial time algorithm for finding a $(1/2+\delta)$-ANE. For the special case of polymatrix graphical games on a tree graph, Barman et al [BLP15] give a PTAS.

**Followup work on PCP for PPAD** Following our work, [BPR16] conjecture the following robust strengthening of our main theorem: given a polymatrix, graphical game, it is PPAD-hard to find a strategy profile where a $(1-\delta)$-fraction of the players are playing $\epsilon$-approximate best response, for some constants $\epsilon, \delta > 0$. (Our Theorem 1 corresponds to the case of $\delta = 1/\mathsf{poly}(n)$.) [BPR16] prove that this so-called "PCP for PPAD" is equivalent to a robust variant of our Theorem 2, and implies robust versions of Corollaries 2 and 5. This conjecture remains an open problem.

[Rub16] uses related ideas to prove that the quasi-polynomial time algorithm of Lipton et al [LMM03] for $\epsilon$-ANE in two-player games is essentially tight: improving to polynomial time would imply an unlikely subexponential-time algorithm for EndOfTheLine.

# 2 Proof overview

We begin our proof of Theorem 1 with the EndOfTheLine problem over $\{0,1\}^n$ [DGP09]. Our first step (Section 4) is to reduce this problem to the problem of finding an approximate fixed point of a particular continuous function, inspired by the work of Hirsch et al [HPV89] (this reduction also appeared in [Rub14]). Our second step (Section 5) is to reduce the problem of finding an approximate fixed point of this particular function to that of finding an approximate solution to a generalized circuit. In Section 6, we further reduce to a generalized circuit with fan-out 2. Our final step (Section 7) is to reduce the problem of finding an approximate solution to a circuit with fan-out 2 to that of finding an approximate Nash equilibrium. This reduction follows directly from the game gadgets of Daskalakis et al [DGP09] which establishes the result for $\epsilon$-WSNE. We complete the proof by pointing out that for bounded degree graphical games, finding an $\epsilon$-WSNE reduces to finding a $\Theta(\epsilon^2)$-ANE.

Our main technical contribution is in the second step, which reduces finding an approximate fixed point of a particular continuous function, to finding an approximate solution to [CDT09]'s generalized circuit problem. We provide further details below (see also Section 5).

## Obtaining a constant hardness of $\epsilon$-GCIRCUIT

A key idea that enables us to improve over previous hardness of approximation for $\epsilon-$GCIRCUIT (and Nash equilibrium) [DGP09, CDT09] is that we start from a better instance of Brouwer function due to [HPV89]. The first advantage is that our instance is simply harder to approximate: finding an $\epsilon$-approximate fixed point (i.e. $\mathbf{x}$ such that $\|f(\mathbf{x}) - \mathbf{x}\|_\infty \leq \epsilon$) is PPAD-hard for $\epsilon = \Omega(1)$ (as opposed to $\epsilon = 1/\exp(n)$ for [DGP09] and $\epsilon = 1/\mathsf{poly}(n)$ in [CDT09]). This advantage of the Hirsch et al construction was used before by Babichenko [Bab14] and also in [Rub14].

**A simpler averaging gadget** Our construction of hard instances of Brouwer functions, as do the ones form previous works, partitions the (continuous) hypercube into subcubes, and define the function separately on each subcube. When we construct a circuit that approximately simulates such a Brouwer function, we have a problem near the facets of the subcubes: using *approximate gates* and *brittle comparators* (both defined formally in Section 3.2), one cannot determine to which subcube the input belongs. This is the most challenging part of our reduction, as was also the case in [DGP09, CDT09].

Originally, Daskalakis et al [DGP09] tackled this obstacle by approximating $f(\mathbf{x})$ as the average over a ball around $\mathbf{x}$. The key observation is that even if $\mathbf{x}$ is close to a facet between subcubes, most of the points in its neighborhoods will be sufficiently far. Yet if $f$ is Lipschitz they are mapped approximately to the same point as $\mathbf{x}$. This works fine in $O(1)$ dimensions, but then the inapproximability parameter is inherently exponentially small (in constant dimensions, it is easy to construct a $1/\mathsf{poly}(n)$-net over the unit hypercube). For $\mathsf{poly}(n)$ dimensions, the (discretization of the) ball around $\mathbf{x}$ contains exponentially many points.

Chen et al [CDT09] overcome this problem using *equiangle sampling*: consider many translations of the input vector by adding small multiples of the all-ones vector; compute the displacement for each translation, and average. Since each translation may be close to a facet in a different dimension, Chen et al consider a polynomial number of translations. Thus, all translations must be polynomially close to each other - otherwise they will be too far to approximate the true input.

We avoid this problem by observing another nice property of the [HPV89]'s construction: when the input vector lies near two or more facets, the displace-

Figure 1: Comparison of averaging gadgets

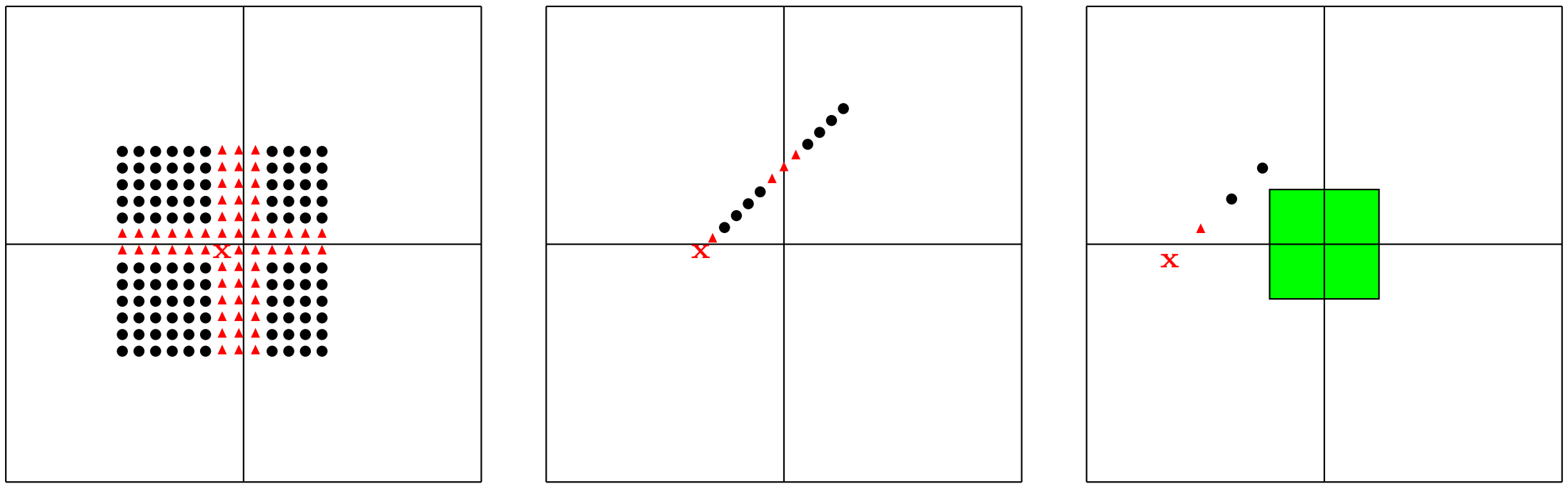

Daskalakis et al [DGP09] Chen et al [CDT09] This paper

A comparison of the averaging gadgets of [DGP09], [CDT09], and this paper. **x** is the point whose displacement we would like to estimate using imprecise gates and brittle comparators. Points that are too close to a facet between subcubes are denoted by triangles, while points that are sufficiently far are denoted by circles. Finally, in this paper we have a "safe" zone (shaded) around the corner where we don't need to parse the subcube; thus we only need to avoid one facet.

ment is (approximately) the same, regardless of the subcube. Once we rule out such points, it suffices to sample only a constant number of points (as at most one of them may be too close to a facet). See also illustration in Figure 2.

**Completing the proof** Given a point $\mathbf{x}' \approx \mathbf{x}$ which is safely in the interior of one subcube, we can parse the corresponding binary vector, use logical operator gates to simulate the EndOfTheLine circuit, and then approximately compute $f\left(\mathbf{x}'\right)$. This is tedious, but mostly straightforward.

One particular challenge that nevertheless arises is preventing the error from accumulating when concatenating approximate gates. Of course this is more difficult in our setting where each gate may err by a constant $\epsilon > 0$. Fortunately, the definition of $\epsilon$-Gcircuit provides logical operator gates that round the output to $\{0, 1\}$ before introducing new error. As long as the inputs are unambiguous bits, approximate logical operator gates can be concatenated without accumulating errors.

In order to carry out the reduction to Nash equilibrium (Section 7), we must first ensure that every gate in our generalized circuit has a constant fan-out (Section 6). We can replace each logical operator gate with a binary tree of fan-out 2, alternating negation gates (that do not accumulate error). Given an arithmetic gate with large fan-out, we convert its output to unary representation[2] using a constant number of (fan-out 2) gates. Then we copy the unary representation using a binary tree of negation gates. Finally, we convert each copy back to a real number using a constant number of gates.

## 2.1 Alternative proofs

We informally sketch a couple of different approaches that could lead to our main theorem or similar results.

### No averaging gadget

The simpler averaging gadget was perhaps the main breakthrough that enabled us to improve on the results in [Rub14]. In hindsight it seems that we could completely avoid the use of averaging gadgets. As we discuss in the previous subsection, reducing from the Hirsch et al Brouwer function allows us to treat *corners*, i.e. the intersection of two or more facets, differently. In particular, near corners we don't need to determine to which subcube our input vector belongs because the displacement is always the same. The simpler averaging gadget is used when we are close to one facet.

Alternatively, one could construct a gadget that determines that the input is close to a particular facet *-without deciding on which side of the facet-* and compute the Hirsch et al displacement on that facet. Thus, no averaging gadget is needed, but some care is necessary when stitching together the corner gadget (which would be the same as this paper), the facet gadget, and the interior point gadget.

### A richer set of gates

There is also a simpler and more intuitive reduction that gives somewhat weaker results, namely: PPAD-completeness for degree 3 graphical games with a constant number of actions per player.

[2]Unary representation of numbers with constant precision is prevalent throughout our implementation of the generalized circuit. We prefer unary representation over binary, because in the former at most one bit can be ambiguous due to the use of brittle comparators.

The most arduous part in our proof is the second step, which takes us from the Hirsch et al Brouwer function to a generalized circuit. In order to prove this reduction, we must design a circuit (or an algorithm) that implements the equiangle sampling and the Hirsch et al Brouwer function using the limited set of gates allowed in Chen et al's definition of $\epsilon$-GCIRCUIT. This part could be simplified using a more expressive set of gates.

A recent reduction that appeared in Eran Shmaya's blog [Shm12] would essentially allow us to to replace the $\epsilon$-GCIRCUIT gates with any gates of bounded fan-in and fan-out that compute $c$-Lipschitz functions for any constant $c$. We briefly sketch here this reduction from an arbitrary generalized circuit to a graphical game:

Each line in the circuit corresponds to a player, and we connect all players that share a common gate. The pure strategies of each player correspond to an $O(\epsilon)$-discretization of $[0,1]$. When the player that corresponds to the output of gate $G$ plays strategy $b$, and the input players play strategies $a_1$ and $a_2$, the utility to the output player is given by

$$u(b, a_1, a_2) = -\left|b - G(a_1, a_2)\right|^2$$

It can be shown that in any $O(\epsilon^2)$-WSNE of this game, the output player only uses the strategy[3] $b^*$ which is closest to $\mathbb{E}_{a_1,a_2}[G(a_1,a_2)]$ (i.e. the expectation of $G(a_1,a_2)$ over the mixed strategies of the input players).

# 3 Preliminaries

Throughout this paper we use the max-norm as the default measure of distance. In particular, when we say that $f$ is $M$-Lipschitz we mean that for every $\mathbf{x}$ and $\mathbf{y}$ in the domain of $f$, $\|f(\mathbf{x}) - f(\mathbf{y})\|_\infty \leq M\|\mathbf{x} - \mathbf{y}\|_\infty$.

A large part of our paper deals with approximate solutions to equations. We adopt the notation of writing $x = y \pm \epsilon$ to imply that $x \in (y - \epsilon, y + \epsilon)$.

We use $\boldsymbol{\xi}_i$ to denote the $i$-th standard basis vector, and $0^n$ to denote the all-zeros vector.

[3] There may be two strategies which are equally close to $\mathbb{E}_{a_1,a_2}[G(a_1,a_2)]$, in which case the output player may use a mixed strategy.

**The EndOfTheLine problem**

Our reduction starts from the EndOfTheLine problem. This problem was implicit in [Pap94], and explicitly defined in [DGP09].

**Definition 1.** EndOfTheLine: ([DGP09]) Given two circuits $S$ and $P$, with $n$ input bits and $n$ output bits each, such that $P(0^n) = 0^n \neq S(0^n)$, find an input $x \in \{0,1\}^n$ such that $P(S(x)) \neq x$ or $S(P(x)) \neq x \neq 0^n$.

We like to interpret EndOfTheLine as a problem over a graph which is implicitly defined by circuits $S$ and $P$: every vertex $x \in \{0,1\}^n$ has one incoming edge from $P(x)$, and one outgoing edge to $S(x)$. The only exceptions are sources, for which $P(x) = x$, and sinks, for which $S(x) = x$. The special vertex $0^n$ has an odd degree (no incoming edge), so by a parity argument, there must be at least one more vertex with an odd degree. The goal is to find such a vertex. For technical reasons, we also allow the solution to point out inconsistencies in the graph definition (i.e. $x$ thinks it has an incoming edge from $y$, but $y$ doesn't have an outgoing edge to $x$).

**Theorem.** *(Essentially [Pap94])* EndOfTheLine *is* PPAD*-complete.*

## 3.1 $\epsilon$-Well Supported Nash Equilibrium vs $\epsilon$-Approximate Nash Equilibrium

A mixed strategy of player $i$ is a distribution over $i$'s set of actions, denoted $x_i \in \Delta A_i$. We say that a vector of mixed strategies $\mathbf{x} \in \times_j \Delta A_j$ is a *Nash equilibrium* if every strategy $a_i$ in the support of $x_i$ is a best response to the actions of the mixed strategies of the rest of the players, $x_{-i}$. Formally,

$$\forall a_i \in \text{Supp}(x_i) \quad \mathbb{E}_{a_{-i}\sim x_{-i}}\left[u_i(a_i, a_{-i})\right] = \max_{a' \in A_i} \mathbb{E}_{a_{-i}\sim x_{-i}}\left[u_i(a', a_{-i})\right] .$$

Equivalently, $\mathbf{x}$ is a Nash equilibrium if each mixed strategy $x_i$ is a best mixed response to $x_{-i}$:

$$\mathbb{E}_{\mathbf{a}\sim\mathbf{x}}\left[u_i(\mathbf{a})\right] = \max_{x_i' \in \Delta A_i} \mathbb{E}_{\mathbf{a}\sim\left(x_i'; x_{-i}\right)}\left[u_i(\mathbf{a})\right] .$$

Each of those equivalent definitions can be generalized to include approximation in a different way. (Of course, there are also other interesting generalizations of Nash equilibria to approximate settings.) We say that $\mathbf{x}$ is an $\epsilon$-*approximate Nash equilibrium* ($\epsilon$-*ANE*) if each $x_i$ is an $\epsilon$-best mixed response to $x_{-i}$:

$$\mathbb{E}_{\mathbf{a}\sim\mathbf{x}}\left[u_i\left(\mathbf{a}\right)\right] \geq \max_{x_i'\in\Delta A_i} \mathbb{E}_{\mathbf{a}\sim\left(x_i';x_{-i}\right)}\left[u_i\left(\mathbf{a}\right)\right] - \epsilon\,.$$

On the other hand, we generalize the first definition of Nash equilibrium by saying that $\mathbf{x}$ is a $\epsilon$-*Well Supported Nash Equilibrium* ($\epsilon$-*WSNE*; sometimes also just $\epsilon$-Nash Equilibrium [Das13]) if each $a_i$ in the support of $x_i$ is an $\epsilon$-best response to $x_{-i}$:

$$\forall a_i \in \operatorname{Supp}\left(x_i\right) \;\; \mathbb{E}_{a_{-i}\sim x_{-i}}\left[u_i\left(a_i, a_{-i}\right)\right] \geq \max_{a'\in A_i} \mathbb{E}_{a_{-i}\sim x_{-i}}\left[u_i\left(a', a_{-i}\right)\right] - \epsilon\,.$$

It is easy to see that every $\epsilon$-WSNE is also an $\epsilon$-ANE, but the converse is false. In Lemma 5 we prove that given an $\epsilon$-ANE in a graphical game with incoming degree $d_{in}$, it is possible to find a $\Theta\left(\sqrt{\epsilon}d_{in}\right)$-WSNE. We note that this approach is unlikely to give a lower bound on the *query complexity* for $\epsilon$-ANE: graphical games of bounded incoming degree can be learned in a polynomial number of queries; furthermore, due to [GR14], for any concisely represented game, it is possible to find an $\epsilon$-WSNE (and hence also $\epsilon$-ANE) in a polynomial number of queries.

## 3.2 The $\epsilon$-Gcircuit problem

Generalized circuits are similar to the standard algebraic circuits, the main difference being that generalized circuits contain cycles, which allow them to verify fixed points of continuous functions. We restrict the class of generalized circuits to include only a particular list of gates described below. Formally,

**Definition 2.** [Generalized circuits, [CDT09]] A *generalized circuit* $\mathcal{S}$ is a pair $(V, \mathcal{T})$, where $V$ is a set of nodes and $\mathcal{T}$ is a collection of gates. Every gate $T \in \mathcal{T}$ is a 5-tuple $T = G\left(\zeta \mid v_1, v_2 \mid v\right)$, in which $G \in \{G_\zeta, G_{\times\zeta}, G_=, G_+, G_-, G_<, G_\vee, G_\wedge, G_\neg\}$ is the type of the gate; $\zeta \in \mathbb{R} \cup \{nil\}$ is a real parameter; $v_1, v_2 \in V \cup \{nil\}$ are the first and second input nodes of the gate; and $v \in V$ is the output node.

The collection $\mathcal{T}$ of gates must satisfy the following important property: For every two gates $T = G\left(\zeta \mid v_1, v_2 \mid v\right)$ and $T' = G'\left(\zeta' \mid v_1', v_2' \mid v'\right)$ in $\mathcal{T}$, $v \neq v'$.

Alternatively, we can think of each gate as a constraint on the values on the incoming and outgoing wires. We are interested in the following constraint satisfaction problem: given a generalized circuit, find an assignment to all the wires that simultaneously satisfies all the gates. When every gate computes a continuous function of the incoming wires (with inputs and output in $[0,1]$), a solution must exist by Brouwer's fixed point theorem.

In particular, we are interested in the approximate version of this CSP, where we must approximately satisfy every constraint.

**Definition 3.** Given a generalized circuit $\mathcal{S} = (V, \mathcal{T})$, we say that an assignment $\mathbf{x}\colon V \to [0,1]$ *$\epsilon$-approximately satisfies* $\mathcal{S}$ if for each of the following gates, $\mathbf{x}$ satisfies the corresponding constraints:

| Gate | Constraint |
|---|---|
| $G_{\zeta}(\alpha \mid\mid a)$ | $\mathbf{x}[a] = \alpha \pm \epsilon$ |
| $G_{\times\zeta}(\alpha \mid a \mid b)$ | $\mathbf{x}[b] = \alpha \cdot \mathbf{x}[a] \pm \epsilon$ |
| $G_{=}(\mid a \mid b)$ | $\mathbf{x}[b] = \mathbf{x}[a] \pm \epsilon$ |
| $G_{+}(\mid a, b \mid c)$ | $\mathbf{x}[c] = \min(\mathbf{x}[a] + \mathbf{x}[b], 1) \pm \epsilon$ |
| $G_{-}(\mid a, b \mid c)$ | $\mathbf{x}[c] = \max(\mathbf{x}[a] - \mathbf{x}[b], 0) \pm \epsilon$ |
| $G_{<}(\mid a, b \mid c)$ | $\mathbf{x}[c] = \begin{cases} 1 \pm \epsilon & \mathbf{x}[a] < \mathbf{x}[b] - \epsilon \\ 0 \pm \epsilon & \mathbf{x}[a] > \mathbf{x}[b] + \epsilon \end{cases}$ |
| $G_{\vee}(\mid a, b \mid c)$ | $\mathbf{x}[c] = \begin{cases} 1 \pm \epsilon & \mathbf{x}[a] = 1 \pm \epsilon \text{ or } \mathbf{x}[b] = 1 \pm \epsilon \\ 0 \pm \epsilon & \mathbf{x}[a] = 0 \pm \epsilon \text{ and } \mathbf{x}[b] = 0 \pm \epsilon \end{cases}$ |
| $G_{\wedge}(\mid a, b \mid c)$ | $\mathbf{x}[c] = \begin{cases} 1 \pm \epsilon & \mathbf{x}[a] = 1 \pm \epsilon \text{ and } \mathbf{x}[b] = 1 \pm \epsilon \\ 0 \pm \epsilon & \mathbf{x}[a] = 0 \pm \epsilon \text{ or } \mathbf{x}[b] = 0 \pm \epsilon \end{cases}$ |
| $G_{\neg}(\mid a \mid b)$ | $\mathbf{x}[b] = \begin{cases} 1 \pm \epsilon & \mathbf{x}[a] = 0 \pm \epsilon \\ 0 \pm \epsilon & \mathbf{x}[a] = 1 \pm \epsilon \end{cases}$ |

(Where $G_{\zeta}$ and $G_{\times\zeta}$ also take a parameter $\alpha \in [0,1]$.)

Given a generalized circuit $\mathcal{S} = (V, \mathcal{T})$, $\epsilon$-GCIRCUIT is the problem of finding an assignment that $\epsilon$-approximately satisfies it.

**Brittle comparators** Intuitively, in order for (approximate) solutions to the circuit problem to correspond to (approximate) equilibria, all our gates should implement continuous (Lipschitz) functions. The gate $G_{<}(\mid a, b \mid c)$, for exam-ple, approximates that the function $c(a,b) = \begin{cases} 1 & a < b \\ 0 & a \geq b \end{cases}$, which is not continuous. To overcome this problem, Daskalakis et al [DGP09] defined the *brittle comparator*: when $a$ is ($\epsilon$-) larger than $b$, it outputs 0; when $b$ is ($\epsilon$-) larger than $a$, it outputs 1. However, when $a$ and $b$ are ($\epsilon$-approximately) equal, its behavior is undefined.

Brittleness introduces difficulties in the transition from continuous to discrete solutions. This challenge is overcome by an averaging gadget, which is described in detail in Section 5.

## 3.3 Max-norm Geometry

As we mentioned earlier, throughout this paper we work with the max-norm. This has some implications that may contradict our geometric intuition. For example: *in a max-norm world, a circle is a square.*

**Max-norm interpolation** Given coordinates $x, y \geq 0$, we define the *max-norm angle*[4] that point $(x, y)$ forms with the $X$-axis (in the $XY$-plane) as

$$\theta_{\max}(x,y) = \frac{y}{x+y}$$

The max-norm angle is useful for interpolation. Given the values of $f\colon [0,1]^n \rightarrow [0,1]^n$ on two neighboring facets of the hypercube, we can extend $f$ to all points of the hypercube by *angular interpolation*: interpolate according to the max-norm angle $\theta_{\max}(x_i, x_j)$ where $x_i$ and $x_j$ are the respective distances from the two facets. When $f$ is defined on two opposite facets, we can simply use *Cartesian interpolation*, which again means to interpolate according to the distance from each facet.

**Max-norm local polar coordinates** Given a point $\mathbf{z} \in \mathbb{R}^n$ we define a new *local max-norm polar coordinate* system around $\mathbf{z}$. Every $\mathbf{x} \in \mathbb{R}^n$ is transformed into $\langle r, \mathbf{p} \rangle_{\mathbf{z}} \in \mathbb{R} \times \mathbb{R}^n$ where $r = \|\mathbf{x} - \mathbf{z}\|$ is the max-norm radius, and $\mathbf{p} = (\mathbf{x} - \mathbf{z})/r$ is the max-norm unit vector that points from $\mathbf{z}$ in the direction of $\mathbf{x}$.

[4] Our max-norm angle was called *unit* in [HPV89].

# 4 Finding an approximate Brouwer fixed point is PPAD-hard

In the first step of the proof, we show that finding an approximate Brouwer fixed point is PPAD-hard. Essentially the same reduction also appeared in [Rub14], and it heavily relies on techniques of Hirsch et al [HPV89].

**Theorem 3.** *(Essentially [Rub14]) There exists a constant $\epsilon > 0$ such that, given an arithmetic circuit that computes an $M$-Lipschitz function $f\colon [0,1]^n \to [0,1]^n$, finding an $\epsilon$-fixed point of $f$ (an $\mathbf{x}$ such that $\|f(\mathbf{x}) - \mathbf{x}\| \leq \epsilon$) is PPAD-hard.*

Theorem 3 by itself does not quite suffice for proving our main theorem. In Section 5 we use the specific properties of our construction, in particular those detailed in Fact 1.

*Proof.* In the first step (Subsection 4.1), we embed the EndOfTheLine problem (over $\{0,1\}^n$) as a collection $H$ of vertex-disjoint paths over the $(2n+1)$-dimensional hypercube graph. Given $H$, our second step (Subsection 4.2) is to construct a continuous mapping $f\colon [0,1]^{2n+2} \to [0,1]^{2n+2}$ whose fixed points correspond to ends of paths in $H$. This step generalizes a construction of Hirsch et al [HPV89] for embedding a single path. □

## 4.1 Embedding the EndOfTheLine graph as paths in $\{0,1\}^{2n+1}$

Our first step in the reduction is to embed an EndOfTheLine graph as vertex-disjoint paths on the $(2n+1)$-dimensional hypercube graph. We first recall that the input to the EndOfTheLine problem is given as two circuits $S$ and $P$, which define a directed graph over $G$ over $\{0,1\}^n$. Given $S$ and $P$, we construct a collection $H$ of vertex-disjoint paths and cycles over the $(2n+1)$-dimensional hypercube graph, such that there is a 1-to-1 correspondence between starting and end points of paths in $H$ and starting and end points of lines in $G$.

In order to construct our embedding we divide the $2n+1$ coordinates as follows: the first $n$ coordinates store the current vertex $\mathbf{u}$, the next $n$ coordinates for the next vertex in the line, $\mathbf{v}$, and finally, the last coordinate $b$ stores a compute-next vs copy bit. When $b = 0$, the path proceeds to update $\mathbf{v} \leftarrow S(\mathbf{u})$, bit-by-bit. When this update is complete, the value of $b$ is changed to 1. Whenever $b = 1$, the path proceeds by copying $\mathbf{u} \leftarrow \mathbf{v}$ bit-by-bit, and then changes that value of $b$ again. Finally, when $\mathbf{u} = \mathbf{v} = S(\mathbf{u})$ and $b = 0$, the path reaches an end point. For example, the edge $\mathbf{x} \to \mathbf{y}$ maps into the path:

$$(\mathbf{x}, \mathbf{x}, 0) \to \cdots \to (\mathbf{x}, \mathbf{y}, 0) \to (\mathbf{x}, \mathbf{y}, 1) \to \cdots \to (\mathbf{y}, \mathbf{y}, 1) \to (\mathbf{y}, \mathbf{y}, 0).$$

Notice that the paths in $H$ do not intersect. Furthermore, given a vector in $\mathbf{p} \in \{0,1\}^{2n+1}$, we can output in polynomial time whether $\mathbf{p}$ belongs to a path in $H$, and if so which are the previous and consecutive vectors in the path. It is therefore PPAD-hard to find a starting or end point of any path in $H$ other than $0^{2n+1}$.

## 4.2 Continuous mapping on $[0,1]^{2n+2}$

In order to construct a hard instance of Brouwer function, we use techniques introduced by Hirsch, Papadimitriou, and Vavasis [HPV89]. [HPV89] showed that the number of deterministic value-oracle queries required to find a Brouwer fixed point is exponential in the number of precision digits, the Lipschitz constant of the function, and -most important to us- the dimension of the domain. Our construction is almost identical, with the exception that [HPV89] embed a single path, whereas we embed $H$ which is a collection of paths and cycles.

The continuous Brouwer function is denoted by $f\colon [0,1]^{2n+2} \to [0,1]^{2n+2}$, while the associated displacement function is denoted by $g(\mathbf{x}) = f(\mathbf{x}) - \mathbf{x}$. The following lemma (proven below) completes the proof of Theorem 3.

**Lemma 1.** *The displacement $g$ satisfies:*

1. *$g$ is $O(1)$-Lipschitz (thus, $f$ is also $O(1)$-Lipschitz)*
2. *$\|g(\mathbf{x})\|_\infty = \Omega(1)$ for every $\mathbf{x}$ that does not correspond to the endpoint of the path*
3. *The value of $g$ at each point $\mathbf{x}$ can be computed in polynomial time using $S$ and $P$.*

### 4.2.1 Overview of the construction

The domain of $f$ is the $2n+2$-dimensional (solid) hypercube. The hypercube is divided into subcubes, of side length $h$ (we fix $h = 1/4$). We define $f$ separately on each subcube such that it agrees on the intersections (no interpolation is needed in this sense).

The last ($(2n+2)$-th) dimension is special; we use "up" (resp. "down") to refer to the positive (negative) $(2n+2)$-th direction. All the action takes place in the second-from-bottom $(2n+1)$-dimensional layer of subcubes; this layer is called the *slice*. Within the slice, we also ignore the subcubes that are near the boundary of the hypercube (those compose the *frame*); we are left with the subcubes in the center of the slice, which we call the *picture*. We identify between the vertices of the $(2n+1)$-dimensional hypercube graph (over which $H$ was defined) and the $2^{2n+1}$ subcubes of the picture.

In [HPV89], a single path is embedded inside the picture; this embedded path is called the *tube*. The *home subcube*, the subcube that corresponds to the beginning of the path, is special: all the flow from all subcubes that do not belong to the tube leads to this subcube. For our purposes, we consider many tubes, corresponding to the paths and cycles of $H$. The home subcube continues to be special in the sense described above, and it corresponds to the vertex $0^{2n+1}$.

Below we define the displacement in the following regions, and argue that it satisfies the desiderata of Lemma 1:

- Inside the picture, but not in any tube;
- inside a tube; and
- outside the picture.

#### 4.2.2 Default displacement

Most of the slice has the same *default displacement*: directly upward, i.e. $g(\mathbf{x}) = \delta \boldsymbol{\xi}_{2n+2}$, where $\boldsymbol{\xi}_{2n+2}$ is the $(2n+2)$-unit vector, and $\delta > 0$ is a small constant. Formally,

**Fact 1.** *$g(\mathbf{x}) = \delta \boldsymbol{\xi}_{2n+2}$, for every $\mathbf{x}$ such that at least one of the following holds:*

1. *$\mathbf{x}$ lies on a corner, i.e. the intersection of two or more facets of a subcube;*
2. *$\mathbf{x}$ lies on an outer facet of a tube subcube, i.e. a facet other than the two facets that continue the path; or*
3. *$\mathbf{x}$ lies in a subcube that does not belong to any tube.*

Intuitively, Property 2 implies that *all subcubes -whether they belong to the tube or not- look the same from the outside* (except for the two facets that continue the path). In particular, the displacement on both sides of each facet is the same; so if the displacement is $O(1)$-Lipschitz on each subcube, it is also $O(1)$-Lipschitz on the entire hypercube.

Property 1, stating that *all corners look the same*, is key to the sampling gadget in Section (5), because it liberates us from having to disambiguate the position of a point near the corners (that is, deciding exactly to which subcube it belongs). This is useful because those points that are close to corners, are precisely the ones that are hard to determine using “noisy” gates.

### 4.2.3 Displacement at a tube

The mapping is defined so that in the center of the tube, the flow goes along the direction of the path; slightly outside the center, the flow points towards the center of the tube; further away from the center, the flow goes *against* the direction of the path; at the outer boundary of the tube, as we previously described, the flows goes upwards.

We first define $g$ on facets. Let $\langle r, \mathbf{p} \rangle_{\mathbf{z}}$ be a point on the facet centered at $\mathbf{z}$, and suppose that the tube enters the subcube through $\mathbf{z}$, advancing in the positive $i$-th coordinate. We define

$$g\left(\langle r, \mathbf{p} \rangle_{\mathbf{z}}\right) = \begin{cases} \delta \boldsymbol{\xi}_i & r = 0 \\ -\delta \mathbf{p} & r = h/8 \\ -\delta \boldsymbol{\xi}_i & r = h/4 \\ \delta \boldsymbol{\xi}_{2n+2} & r = h/2 \end{cases} \tag{1}$$

(Recall that $h$ is the subcube side length, and $\delta$ is some small constant.) Notice that at each $r$, the displacement $g$ is $O(1)$-Lipschitz and has magnitude $\|g(\mathbf{x})\|_{\infty} = \Omega(1)$ (thus satisfying the first two desiderata of Lemma 1).

For $r \in (0, h/8)$, interpolate between $\delta \boldsymbol{\xi}_i$ and $-\delta \mathbf{p}$ ([HPV89] call this *radial interpolation*), and similarly for $r \in (h/8, h/4)$ and $r \in (h/4, h/2)$. See also illustration in Figure 2. It is easy to see that the $O(1)$-Lipschitz property is preserved. Notice also that $\boldsymbol{\xi}_i$ is orthogonal to $\mathbf{p}$ and $\boldsymbol{\xi}_{2n+2}$; this guarantees that the interpolation does not lead to cancellation, i.e. we still have $\|g(\mathbf{x})\|_{\infty} = \Omega(1)$.

In the last couple of paragraphs we defined $g$ on two facets for each subcube that belongs to the tubes; for all other points in the tubes we interpolate (angular interpolation) between those two facets: Consider a point $\mathbf{x}$ in the tube, and assume (w.l.o.g.) that $x_i, x_j > 1/2$, and suppose that the value of $f(\cdot)$ on the $y_i = 1/2$ and $y_j = 1/2$ facets of the subcube containing $\mathbf{x}$ is determined by (1). Let

$$\begin{aligned} \mathbf{x}^i &= \left(\mathbf{x}_{-i,j}, \frac{1}{2}, \max\{x_i, x_j\}\right) \\ \mathbf{x}^j &= \left(\mathbf{x}_{-i,j}, \max\{x_i, x_j\}, \frac{1}{2}\right) \end{aligned}$$

denote the corresponding "max-norm projections" to the respective $y_i = 1/2$ and $y_j = 1/2$ facets. We set

$$g(\mathbf{x}) = \theta_{\max}\left(x_i - \frac{1}{2}, x_j - \frac{1}{2}\right) g\left(\mathbf{x}^i\right) + \left(1 - \theta_{\max}\left(x_i - \frac{1}{2}, x_j - \frac{1}{2}\right)\right) g\left(\mathbf{x}^j\right).$$

Notice that $\mathbf{x}^i$ and $\mathbf{x}^j$ are at the same distance from the respective facet centers, i.e. they have they correspond to the same $r$. For each case of (1), the $(i, j)$-components of the displacements at $\mathbf{x}^i$ and $\mathbf{x}^j$ are orthogonal, and for the rest of the components they are aligned. Therefore, when we interpolate between $g\left(\mathbf{x}^i\right)$ and $g\left(\mathbf{x}^j\right)$ there is again no cancellation, i.e. $\|g(\mathbf{x})\|_\infty = \Omega\left(\|g\left(\mathbf{x}^i\right)\|_\infty\right) = \Omega(1)$. Finally, recall that the displacement on each facet is $O(1)$-Lipschitz, and the displacements agree on the intersection of the facets. Therefore the interpolated displacement is $O(1)$-Lipschitz over the entire subcube by a triangle-inequality argument.

The home subcube is defined using (1) as if the tube enters from above, i.e. coming down the $(2n+2)$-dimension, and exits through another facet (in one of the first $(2n+1)$ dimensions) in the direction of the path (here again we have $\|g(\mathbf{x})\|_\infty = \Omega(1)$). For all other starting and end points, we define $g(\mathbf{x}) = \delta\boldsymbol{\xi}_{2n+2}$ on the facet opposite the one that continues the tube, and interpolate between the opposite facets using Cartesian interpolation. Notice that this gives a fixed point when the interpolation cancels the default displacement at the opposite facet, with the displacement $-\delta\boldsymbol{\xi}_{2n+2}$ at the point on the tube facet which is at distance $h/8$ above the path.

Figure 2: A facet of the Hirsch et al construction

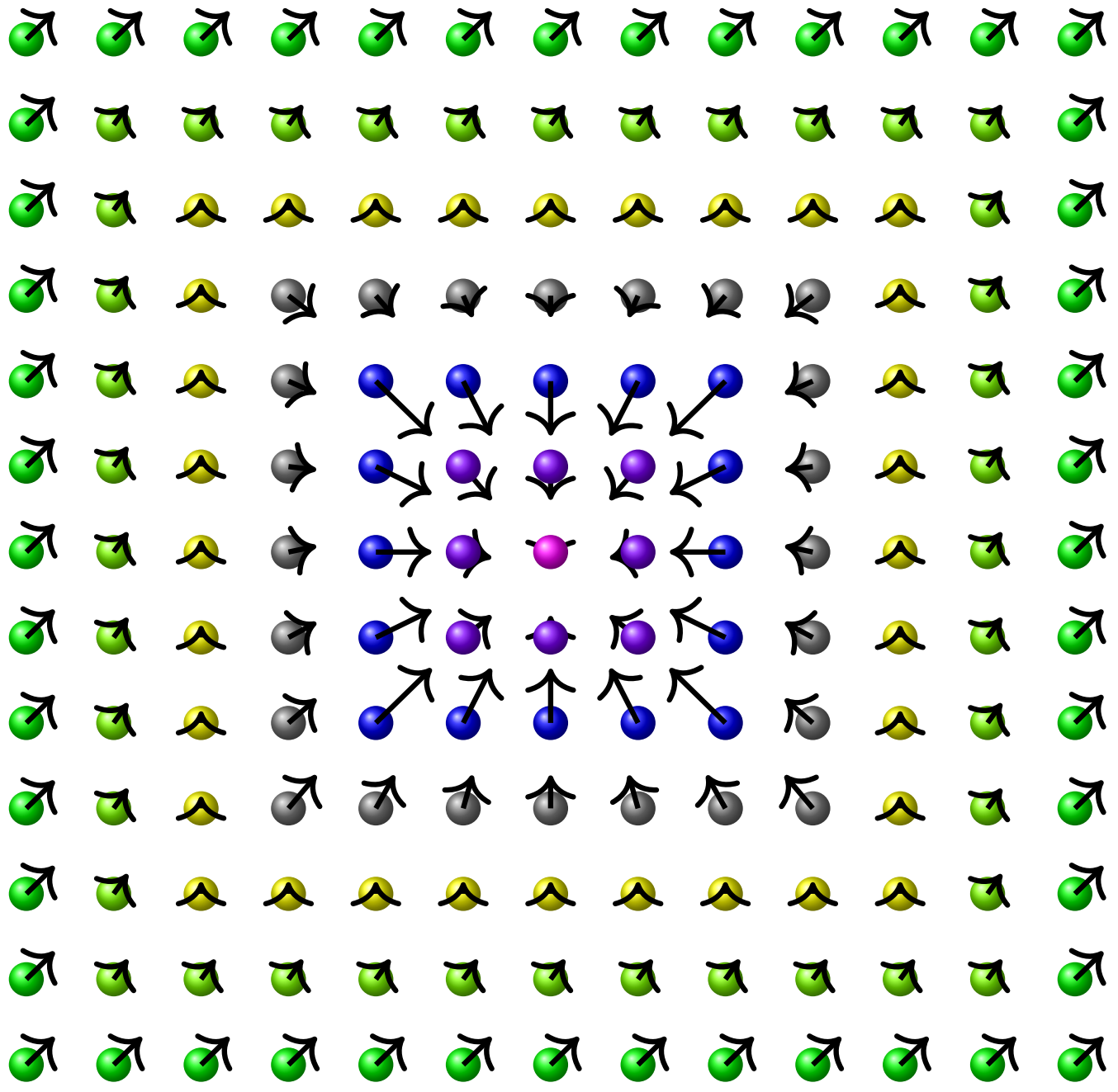

An illustration of the displacement on a facet between two subcubes in a tube; the direction of the path is into the paper. In the center, the displacement points into the paper; slightly further outside, the displacement points towards the center; further outside, the displacement points out of the paper; finally in the outer layer, the displacements points in the special $2n + 2$ dimension.

#### 4.2.4 Outside the picture

For all points in the frame and below the slice, the displacement points directly upward, i.e. $g(\mathbf{x}) = \delta \boldsymbol{\xi}_{2n+2}$. Moving above the slice, let $\mathbf{z}[\text{top}]$ be the point on the top facet of the hypercube which is directly above the center of the home subcube. For all points $\langle r, \mathbf{p} \rangle_{\mathbf{z}[\text{top}]}$ on the top facet of the hypercube, define the displacement as follows:

$$g\left(\langle r, \mathbf{p} \rangle_{\mathbf{z}[\text{top}]}\right) = \begin{cases} -\delta \boldsymbol{\xi}_{2n+2} & r = 0 \\ -\delta \mathbf{p} & r \geq h/8 \end{cases}$$

and interpolate for $r \in (0, h/8)$. Notice that this displacement is $O(1)$-Lipschitz and has $\Omega(1)$ magnitude for each $r$, and this is preserved after interpolation.

Notice that the definition of $g$ on the slice from the previous subsection, implies that all the points in the top facet of the slice, except for the top of the home subcube, point directly upwards. Let $\mathbf{z}[\text{home}]$ denote the center of the top facet of the home subcube. We therefore have that for any $\langle r, \mathbf{p} \rangle_{\mathbf{z}[\text{home}]}$ in the top facet of the slice,

$$g\left(\langle r, \mathbf{p} \rangle_{\mathbf{z}[\text{home}]}\right) = \begin{cases} -\delta \boldsymbol{\xi}_{2n+2} & r = 0 \\ -\delta \mathbf{p} & r = h/8 \\ \delta \boldsymbol{\xi}_{2n+2} & r \geq h/4 \end{cases}$$

where we again interpolate radially for $r$ in $(0, h/8)$ and $(h/8, h/4)$.

Finally, to complete the construction above the slice, simply interpolate (using Cartesian interpolation) between the top facets of the slice and the hypercube. See also illustration in Figure 3.

# 5 From Brouwer to $\epsilon$-GCIRCUIT

**Proposition 1.** *There exists a constant $\epsilon > 0$ such that $\epsilon$-*GCIRCUIT *is* PPAD*-complete.*

*Proof.* We continue to denote the hard Brouwer function by $f\colon [0,1]^{2n+2} \to [0,1]^{2n+2}$, and its associated displacement by $g(\mathbf{y}) = f(\mathbf{y}) - \mathbf{y}$. We design a generalized circuit $\mathcal{S}$ that computes $f$, and verifies that the output is equal to

Figure 3: Outside the picture

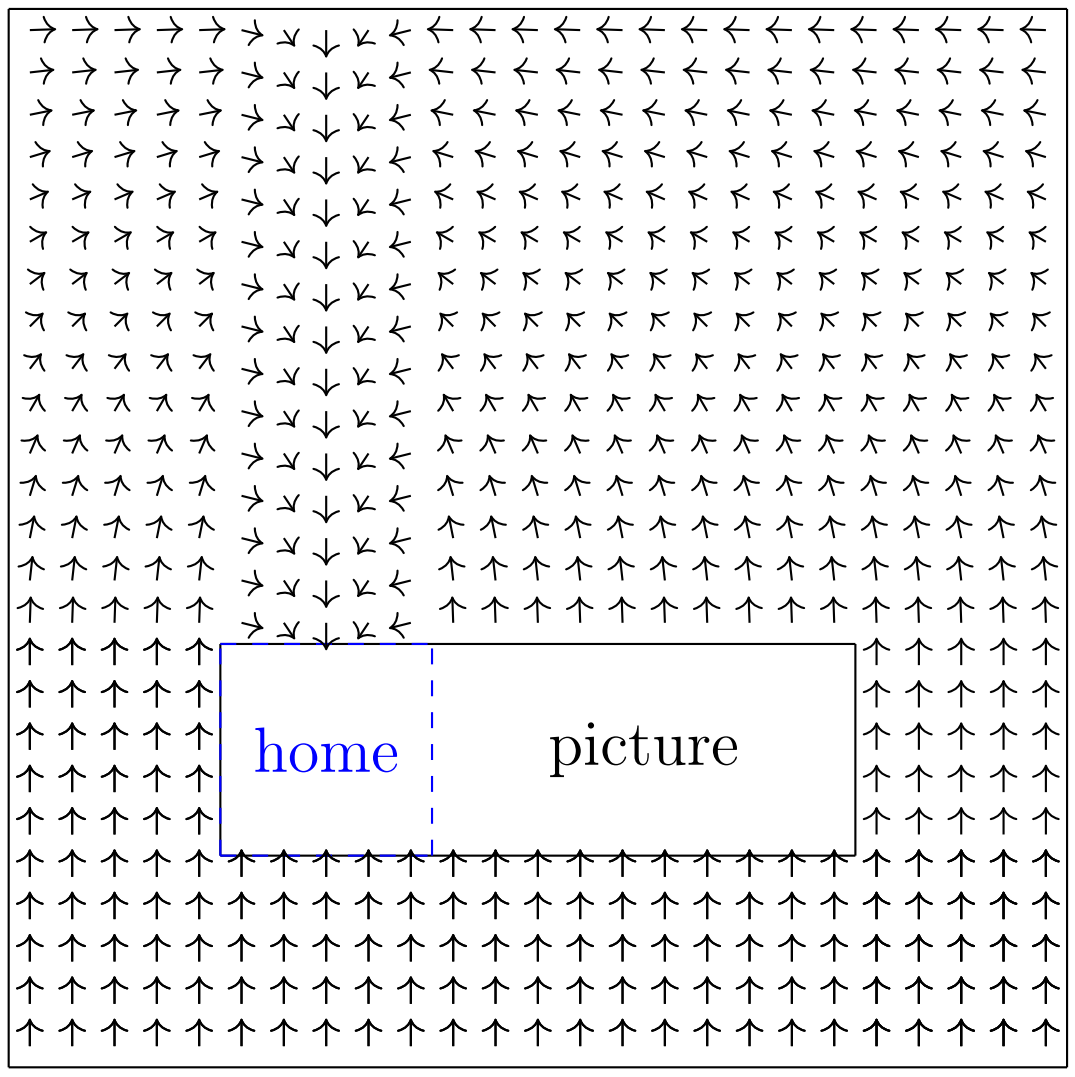

An illustration of the displacement outside the picture.

the input. We show that every $\epsilon$-approximate solution to $\mathcal{S}$ corresponds to an $O\left(\epsilon^{1/4}\right)$-approximate fixed point of $f$.

Recall that the construction from Section (4) divides the hypercube into equal-sized subcubes (of length 1/4). Furthermore, all the paths in $H$ are embedded in the $2^{2n+1}$ subcubes that belong to the picture. For ease of exposition, we present a construction that only works for points in the picture, i.e. $\mathbf{y} \in [1/4, 3/4]^{2n+1} \times [1/4, 1/2]$. It is straightforward how to use the same ideas to extend the circuit to deal with all $\mathbf{y} \in [0,1]^{2n+2}$.

The most challenging part of the construction is the extraction of the information about the local subcube: is it part of a tube? if so, which are the entrance and exit facets? This is done by extracting the binary representation of the current subcube, and feeding it to the (Boolean) circuit that computes $H$ (recall that $H$ is our collection of paths and cycles from Section (4.1)). Notice that whenever we have valid logic bits, i.e. $\mathbf{x}[b] < \epsilon$ or $\mathbf{x}[b] > 1 - \epsilon$, we can perform logic operations on them without increasing the error.

Once we know the behavior of the path on the current subcube, we simply have to locally implement the mapping from the previous section, for which we have a closed form description, using the available gates in the definition of generalized circuits. Since this definition does not include multiplication and division, we implement multiplication and division in Algorithms 2 and 3 in Subsection 5.1.

Our construction has four parts: (1) equiangle sampling segment, (2) computing the displacement, (3) summing the displacement vectors, and (4) closing the loop. The first part contains a new idea introduced in this paper: using a constant size sample. The second part is a more technical but straightforward description of the implementation of the closed-form mapping by approximate gates. The third and fourth parts are essentially identical to [CDT09]. □

## 5.1 Subroutines

In this subsection we show how to implement a few useful subroutines using the gates in the definition of $\epsilon$-GCIRCUIT.

### 5.1.1 If-Else

We begin by describing how to implement a simple if-else. Similar ideas can be used to implement more involved cases such as (1).

*Claim* 1. In any $\epsilon$-approximate solution to IF-ELSE($\mid a, b, c \mid d$),

$$\mathbf{x}[d] = \begin{cases} \mathbf{x}[c] \pm O(\epsilon) & \text{if } \mathbf{x}[a] < \sqrt{\epsilon} \\ \mathbf{x}[b] \pm O(\epsilon) & \text{if } \mathbf{x}[a] > 1 - \sqrt{\epsilon} \end{cases}.$$

**Algorithm 1** IF-ELSE($\mid a, b, c \mid d$)

1. $G_{\neg}(\mid a \mid \overline{a})$# $\overline{a}$ is the negation of $a$
2. $G_{-}(\mid b, \overline{a} \mid b')$# $b'$ is (approximately) equal to $b$ iff $a = 1$
3. $G_{\neg}\left(\mid \overline{a} \mid \overline{\overline{a}}\right)$# $\overline{\overline{a}}$ is the roudning of $a$ to $\{0, 1\}$
4. $G_{-}\left(\mid c, \overline{\overline{a}} \mid c'\right)$# $c'$ is (approximately) equal to $c$ iff $a = 0$
5. $G_{-}(\mid b', c' \mid d)$

*Proof.* By definition of $G_\neg$, we have that

$$\mathbf{x}[\overline{a}] = \begin{cases} 1 \pm \epsilon & \text{if } \mathbf{x}[a] < \sqrt{\epsilon} \\ 0 \pm \epsilon & \text{if } \mathbf{x}[a] > 1 - \sqrt{\epsilon} \end{cases}.$$

Therefore by definition of $G_-$,

$$\mathbf{x}[b'] = \begin{cases} 0 \pm O(\epsilon) & \text{if } \mathbf{x}[a] < \sqrt{\epsilon} \\ \mathbf{x}[b] \pm O(\epsilon) & \text{if } \mathbf{x}[a] > 1 - \sqrt{\epsilon} \end{cases}.$$

Similarly,

$$\mathbf{x}[c'] = \begin{cases} \mathbf{x}[c] \pm O(\epsilon) & \text{if } \mathbf{x}[a] < \sqrt{\epsilon} \\ 0 \pm O(\epsilon) & \text{if } \mathbf{x}[a] > 1 - \sqrt{\epsilon} \end{cases}.$$

Finally, the claim follows by definition of $G_+$. □

### 5.1.2 Multiply

*Claim* 2. In any $\epsilon$-approximate solution to MULTIPLY($\mid a, b \mid c$),

$$\mathbf{x}[c] = \mathbf{x}[a] \cdot \mathbf{x}[b] \pm O\left(\sqrt{\epsilon}\right).$$

*Proof.* For any $k$, the first gate implies that

$$\mathbf{x}[\zeta_k] = k\sqrt{\epsilon} \pm \epsilon.$$

The second gate thus gives

$$\mathbf{x}[\overline{a_k}] = \begin{cases} 0 \pm \epsilon & \text{if } \mathbf{x}[a] > k\sqrt{\epsilon} + O(\epsilon) \\ 1 \pm \epsilon & \text{if } \mathbf{x}[a] < k\sqrt{\epsilon} - O(\epsilon). \end{cases} \tag{2}$$

Notice that the above equation is ambiguous for at most one value of $k$. In particular,

$$\sum_k (1 - \mathbf{x}[\overline{a_k}])\sqrt{\epsilon} = \mathbf{x}[a] \pm O\left(\sqrt{\epsilon}\right). \tag{3}$$

We also have

$$\mathbf{x}[d_k] = \mathbf{x}[b] \cdot \sqrt{\epsilon} \pm \epsilon.$$

**Algorithm 2** MULTIPLY($| \, a, b \mid c$)

1. $G_\zeta \left(0 \mid\mid h_0\right)$

2. **for** each $k \in [1/\sqrt{\epsilon}]$:

   (a) $G_\zeta \left(k\sqrt{\epsilon} \mid\mid \zeta_k\right)$, $G_< \left(\mid a, \zeta_k \mid \overline{a_k}\right)$
   # The vector $(\overline{a_k})$ is the unary representation of $a$

   $$\sum_{k\colon \mathrm{X}[\overline{a_k}]<\epsilon} \sqrt{\epsilon} = \max_{k\colon \mathrm{X}[\overline{a_k}]<\epsilon} k\sqrt{\epsilon} = \mathbf{x}\left[a\right] \pm O\left(\sqrt{\epsilon}\right)$$

   (b) $G_{\times\zeta} \left(\sqrt{\epsilon} \mid b \mid d_k\right)$
   # The vector $(d_k)$ is simply equal to $b \cdot \sqrt{\epsilon}$ everywhere.

   $$\sum_{k\colon \mathrm{X}[\overline{a_k}]<\epsilon} \mathbf{x}\left[d_k\right] = \left(\sum_{k\colon \mathrm{X}[\overline{a_k}]<1-\epsilon} \sqrt{\epsilon}\right) \cdot \mathbf{x}\left[b\right] \pm O\left(\sqrt{\epsilon}\right) = \mathbf{x}\left[a\right] \cdot \mathbf{x}\left[b\right] \pm O\left(\sqrt{\epsilon}\right)$$

   (c) $G_- \left(\mid d_k, \overline{a_k} \mid e_k\right)$
   # The vector $(e_k)$ is $b \cdot \sqrt{\epsilon}$ only when $(\overline{a_k}) < \epsilon$.

   $$\sum_{k\colon \mathrm{X}[\overline{a_k}]<\epsilon} \mathbf{x}\left[e_k\right] = \mathbf{x}\left[a\right] \cdot \mathbf{x}\left[b\right] \pm O\left(\sqrt{\epsilon}\right)$$

   (d) $G_+ \left(\mid h_{k-1}, e_k \mid h_k\right)$
   # Finally, we sum the $e_k$'s to get $a \cdot b$

   $$\mathbf{x}\left[h_{1/\sqrt{\epsilon}}\right] = \mathbf{x}\left[a\right] \cdot \mathbf{x}\left[b\right] \pm O\left(\sqrt{\epsilon}\right)$$

3. $G_= \left(\mid h_{1/\sqrt{\epsilon}} \mid c\right)$

The subtraction gate zeros $\mathbf{x}[d_k]$ for all $k$ such that $\mathbf{x}[a] < k\sqrt{\epsilon} - O(\epsilon)$, and has negligible effect for $k$ such that $\mathbf{x}[a] > k\sqrt{\epsilon} + O(\epsilon)$:

$$\mathbf{x}[e_k] = \begin{cases} \mathbf{x}[b] \cdot \sqrt{\epsilon} \pm 2\epsilon & \text{if } \mathbf{x}[a] > k\sqrt{\epsilon} + O(\epsilon) \\ 0 \pm 2\epsilon & \text{if } \mathbf{x}[a] < k\sqrt{\epsilon} - O(\epsilon). \end{cases}$$

The sum of the $\mathbf{x}[e_k]$'s satisfies:

$$\sum_k \mathbf{x}[e_k] = \mathbf{x}[a] \cdot \mathbf{x}[b] \pm O\left(\sqrt{\epsilon}\right),$$

where we have an error of $\pm O(\sqrt{\epsilon})$ arising from aggregating $\pm 2\epsilon$ for $1/\sqrt{\epsilon}$ distinct $k$'s, and another $\pm O(\sqrt{\epsilon})$ from (3).

By induction, each $h_k$ is approximately equal to the sum of the first $\mathbf{x}[e_j]$'s:

$$\mathbf{x}[h_k] = \sum_{j=1}^{k} \mathbf{x}[e_j] \pm k\epsilon.$$

In particular, we have

$$\begin{aligned} \mathbf{x}\left[h_{1/\sqrt{\epsilon}}\right] &= \sum_k \mathbf{x}[e_k] \pm \sqrt{\epsilon} \\ &= \mathbf{x}[a] \cdot \mathbf{x}[b] \pm O\left(\sqrt{\epsilon}\right). \end{aligned}$$

□

### 5.1.3 Divide

*Claim* 3. In any $\epsilon$-approximate solution to DIVIDE($\mid a, b \mid c$),

$$\mathbf{x}[c] \cdot \mathbf{x}[b] = \mathbf{x}[a] \pm O\left(\sqrt{\epsilon}\right).$$

Notice that for Algorithm 3, in any $\epsilon$-approximate solution, $\mathbf{x}[c] = \mathbf{x}[a] / \mathbf{x}[b] \pm O(\sqrt{\epsilon}) / \mathbf{x}[b]$; when $\mathbf{x}[b]$ and $\epsilon$ are bounded away from 0, this is only a constant factor increase in the error.

*Proof.* For each $k$, we have

$$\mathbf{x}[b_k] = k\sqrt{\epsilon} \cdot \mathbf{x}[b] \pm \epsilon.$$

**Algorithm 3** DIVIDE($\mid a, b \mid c$)

1. $G_\zeta\left(0 \mid\mid h_0\right)$
2. **for** each $k \in [1/\sqrt{\epsilon}]$:

   (a) $G_{\times\zeta}\left(k\sqrt{\epsilon} \mid b \mid b_k\right)$, $G_<\left(\mid b_k, a \mid d_k\right)$
   # The vector $(d_k)$ is the unary representation of $a/b$

   $$\left(\sum_{k\colon \mathrm{X}[d_k]>\epsilon} \sqrt{\epsilon}\right) \cdot \mathbf{x}[b] = \left(\max_{k\colon \mathrm{X}[d_k]>\epsilon} k\sqrt{\epsilon}\right) \cdot \mathbf{x}[b] = \mathbf{x}[a] \pm O\left(\sqrt{\epsilon}\right)$$

   (b) $G_{\times\zeta}\left(\sqrt{\epsilon} \mid d_k \mid e_k\right)$
   # The vector $(e_k)$ is a $(\sqrt{\epsilon})$-scaled version of $(d_k)$

   $$\left(\sum \mathbf{x}[e_k]\right) \cdot \mathbf{x}[b] = \mathbf{x}[a] \pm O\left(\sqrt{\epsilon}\right)$$

   (c) $G_+\left(\mid h_{k-1}, e_k \mid h_k\right)$
   # Finally, we sum the $e_k$'s

   $$\mathbf{x}\left[h_{1/\sqrt{\epsilon}}\right] \cdot \mathbf{x}[b] = \mathbf{x}[a] \pm O\left(\sqrt{\epsilon}\right)$$

3. $G_=\left(\mid h_{1/\sqrt{\epsilon}} \mid c\right)$

Thus also

$$\mathbf{x}[d_k] = \begin{cases} 1 \pm \epsilon & \text{if}\, \mathbf{x}[a] > k\sqrt{\epsilon} \cdot \mathbf{x}[b] + O(\epsilon) \\ 0 \pm \epsilon & \text{if}\, \mathbf{x}[a] < k\sqrt{\epsilon} \cdot \mathbf{x}[b] - O(\epsilon). \end{cases}$$

Notice that $\mathbf{x}[d_k]$ is ambiguous for at most $k$. Furthermore, aggregating the $\pm\epsilon$ error over $1/\sqrt{\epsilon}$ distinct $k$'s, we have:

$$\sum \mathbf{x}[d_k] \cdot \sqrt{\epsilon}\mathbf{x}[b] = \mathbf{x}[a] \pm O\left(\sqrt{\epsilon}\right).$$

$\mathbf{x}[e_k]$'s are a step closer to what we need:

$$\mathbf{x}[e_k] = \mathbf{x}[d_k]\sqrt{\epsilon} \pm \epsilon,$$

and therefore also

$$\sum \mathbf{x}[e_k] \cdot \mathbf{x}[b] = \mathbf{x}[a] \pm O\left(\sqrt{\epsilon}\right). \tag{4}$$

Finally, by induction

$$\mathbf{x}\left[h_{1/\sqrt{\epsilon}}\right] = \sum \mathbf{x}[e_k] \pm \sqrt{\epsilon},$$

and the claim follows by plugging into (4). □

#### 5.1.4 Max

*Claim* 4. In any $\epsilon$-approximate solution to $\text{MAX}(\mid a_1, \ldots a_n \mid b)$,

$$\mathbf{x}[b] = \max \mathbf{x}[a_i] \pm O\left(\sqrt{\epsilon}\right).$$

*Proof.* Similarly to (2), we have that for each $i, k$

$$\mathbf{x}[c_{k,i}] = \begin{cases} 0 \pm \epsilon & \text{if}\, \mathbf{x}[a_i] < k\sqrt{\epsilon} - O(\epsilon) \\ 1 \pm \epsilon & \text{if}\, \mathbf{x}[a_i] > k\sqrt{\epsilon} + O(\epsilon). \end{cases}$$

For each $k$, taking OR of all the $c_{k,i}$'s gives (approximately) 1 iff any of the $\mathbf{x}[a_i]$'s is sufficiently large; in particular if the maximum is:

$$\mathbf{x}[d_{k,n}] = \begin{cases} 0 \pm \epsilon & \text{if } \max_i \mathbf{x}[a_i] < k\sqrt{\epsilon} - O(\epsilon) \\ 1 \pm \epsilon & \text{if } \max_i \mathbf{x}[a_i] > k\sqrt{\epsilon} + O(\epsilon). \end{cases}$$

**Algorithm 4** MAX($\mid a_1, \ldots a_n \mid b$)

1. $G_\zeta\left(0 \,||\, h_0\right)$
2. **for** each $k \in [1/\sqrt{\epsilon}]$:
   (a) $G_\zeta\left(k\sqrt{\epsilon} \,||\, \zeta_k\right)$
   (b) $G_\zeta\left(0 \,||\, d_{k,0}\right)$
   (c) **for** each $i \in [n]$:
      i. $G_<\left(\mid \zeta_k, a_i \mid c_{k,i}\right)$
         # The vector $\left(c_{k,i}\right)_k$ is the unary representation of $a_i$:
         $$\forall i \quad \left(\max_{k:\, \mathrm{x}\left[c_{k,i}\right]>\epsilon} k\sqrt{\epsilon}\right) = \mathbf{x}\left[a_i\right] \pm O\left(\sqrt{\epsilon}\right)$$
      ii. $G_\vee\left(\mid d_{k,i-1}, c_{k,i} \mid d_{k,i}\right)$
      # The vector $\left(d_{k,n}\right)$ is the unary representation of $\max a_i$:
      $$\left(\max_{k:\, \mathrm{x}\left[d_{k,n}\right]>\epsilon} k\sqrt{\epsilon}\right) = \max \mathbf{x}\left[a_i\right] \pm O\left(\sqrt{\epsilon}\right)$$
   (d) $G_{\times\zeta}\left(\sqrt{\epsilon} \mid d_{k,n} \mid e_k\right)$
      # The vector $\left(e_k\right)$ is a $\left(\sqrt{\epsilon}\right)$-scaled version of $\left(d_k\right)$
      $$\left(\sum \mathbf{x}\left[e_k\right]\right) = \max \mathbf{x}\left[a_i\right] \pm O\left(\sqrt{\epsilon}\right)$$
   (e) $G_+\left(\mid h_{k-1}, e_k \mid h_k\right)$
      # Finally, we sum the $e_k$'s
      $$\mathbf{x}\left[h_{1/\sqrt{\epsilon}}\right] = \max \mathbf{x}\left[a_i\right] \pm O\left(\sqrt{\epsilon}\right)$$
3. $G_=\left(\mid h_{1/\sqrt{\epsilon}} \mid b\right)$

Therefore also (similarly to (3)),

$$\sum_k \mathbf{x}\left[d_{k,n}\right]\sqrt{\epsilon} = \max_i \mathbf{x}\left[a_i\right] \pm O\left(\sqrt{\epsilon}\right).$$

The $\mathbf{x}\left[e_k\right]$ take care of scaling by $\sqrt{\epsilon}$:

$$\sum_k \mathbf{x}\left[e_k\right] = \max_i \mathbf{x}\left[a_i\right] \pm O\left(\sqrt{\epsilon}\right).$$

Finally, by induction,

$$\mathbf{x}\left[h_{1/\sqrt{\epsilon}}\right] = \max \mathbf{x}\left[a_i\right] \pm O\left(\sqrt{\epsilon}\right).$$

□

#### 5.1.5 Interpolate

*Claim* 5. In any $\epsilon$-approximate solution to INTERPOLATE$(a, w_a, b, w_b \mid c)$,

$$\mathbf{x}[c]\left(\mathbf{x}\left[w_a\right] + \mathbf{x}\left[w_b\right]\right) = \left(\mathbf{x}\left[w_a\right] \cdot \mathbf{x}[a] + \mathbf{x}\left[w_b\right] \cdot \mathbf{x}[b]\right) \pm O\left(\sqrt{\epsilon}\right).$$

*Proof.* By Claim 3, we have

$$\begin{aligned} \mathbf{x}\left[\overline{w_a}\right] \cdot \left(\mathbf{x}\left[w_a\right] + \mathbf{x}\left[w_b\right]\right) &= \mathbf{x}\left[w_a\right] \pm O\left(\sqrt{\epsilon}\right) \\ \mathbf{x}\left[\overline{w_b}\right] \cdot \left(\mathbf{x}\left[w_a\right] + \mathbf{x}\left[w_b\right]\right) &= \mathbf{x}\left[w_b\right] \pm O\left(\sqrt{\epsilon}\right). \end{aligned}$$

Therefore, by Claim 2,

$$\begin{aligned} \mathbf{x}\left[c_a\right] \cdot \left(\mathbf{x}\left[w_a\right] + \mathbf{x}\left[w_b\right]\right) &= \mathbf{x}\left[w_a\right] \cdot \mathbf{x}[c] \pm O\left(\sqrt{\epsilon}\right) \\ \mathbf{x}\left[c_b\right] \cdot \left(\mathbf{x}\left[w_a\right] + \mathbf{x}\left[w_b\right]\right) &= \mathbf{x}\left[w_b\right] \cdot \mathbf{x}[c] \pm O\left(\sqrt{\epsilon}\right). \end{aligned}$$

The claim follows by definition of $G_+$. □

**Algorithm 5** INTERPOLATE$(a, w_a, b, w_b \mid c)$

1. $G_{\times\zeta}\left(1/2 \mid w_a \mid w_{a/2}\right)$ and $G_{\times\zeta}\left(1/2 \mid w_b \mid w_{b/2}\right)$

   # We divide by 2 before adding in order to stay in $[0,1]$

2. $G_{+}\left(\mid w_{a/2}, w_{b/2} \mid w_{a/2+b/2}\right)$ Add the weights

3. DIVIDE$\left(\mid w_{a/2}, w_{a/2+b/2} \mid \overline{w_a}\right)$ and DIVIDE$\left(\mid w_{b/2}, w_{a/2+b/2} \mid \overline{w_b}\right)$,

   # $\overline{w_a}$ and $\overline{w_b}$ are the normalized weights:

$$\mathbf{x}\left[\overline{w_a}\right] \cdot \left(\mathbf{x}\left[w_a\right] + \mathbf{x}\left[w_b\right]\right) = \mathbf{x}\left[w_a\right] \pm O\left(\sqrt{\epsilon}\right)$$

4. MULTIPLY$(\mid \overline{w_a}, a \mid c_a)$ and MULTIPLY$(\mid \overline{w_b}, b \mid c_b)$

   # $c_a$ and $c_b$ are the $a$ and $b$ components, respectively, of $c$:

$$\mathbf{x}\left[c_a\right] = \mathbf{x}\left[w_a\right] \cdot \mathbf{x}\left[a\right] / \left(\mathbf{x}\left[w_a\right] + \mathbf{x}\left[w_b\right]\right) \pm O\left(\sqrt{\epsilon}\right) / \left(\mathbf{x}\left[w_a\right] + \mathbf{x}\left[w_b\right]\right)$$

5. $G_{+}\left(\mid c_a, c_b \mid c\right)$

# Finally, $c$ is the interpolation of $a$ and $b$:

$$\begin{aligned} \mathbf{x}\left[c\right] &= \left(\mathbf{x}\left[w_a\right] \cdot \mathbf{x}\left[a\right] + \mathbf{x}\left[w_b\right] \cdot \mathbf{x}\left[b\right]\right) / \left(\mathbf{x}\left[w_a\right] + \mathbf{x}\left[w_b\right]\right) \\ &\pm O\left(\sqrt{\epsilon}\right) / \left(\mathbf{x}\left[w_a\right] + \mathbf{x}\left[w_b\right]\right) \end{aligned}$$

## 5.2 Equiangle sampling segment

The first information we require in order to compute the Hirsch et al mapping $f(\mathbf{y})$ is about the subcube to which $\mathbf{y}$ belongs: is it part of the tube? if so, which are the entrance and exit facets? In order to answer those questions, we extract the binary representation of the cube. Recall that our circuit uses brittle comparators; thus when $\mathbf{y}$ is close to a facet between subcubes, the behaviour of the brittle comparators may be unpredictable. We start with the easy case, where $\mathbf{y}$ is actually far from every facet:

**Definition 4.** We say that $\mathbf{y}$ is an *interior point* if for every $i$, $|y_i - 1/2| > \epsilon$; otherwise, we say that $\mathbf{y}$ is a *boundary point.*

A very nice property of the Hirsch et al construction is that whenever $\mathbf{y}$ is at the intersection of two or more facets, the displacement is the same: $g(\mathbf{y}) = \delta\boldsymbol{\xi}_{2n+2}$. Thus, by the Lipschitz property of $g$, whenever $\mathbf{y}$ is close to the intersection of two or more facets, the displacement is approximately $\delta\boldsymbol{\xi}_{2n+2}$. For such $\mathbf{y}$'s, we don't care to which subcube they belong.

**Definition 5.** We say that $\mathbf{y}$ is a *corner point* if there exist distinct $i, j \in [2n+2]$ such that $|y_i - 1/2| < \epsilon^{1/4}$ and $|y_j - 1/2| < \epsilon^{1/4}$.

(Notice that $\mathbf{y}$ may be an interior point and a corner point at the same time.)

We still have a hard time handling $\mathbf{y}$'s which are neither an interior point nor a corner point. To mitigate the effect of such $\mathbf{y}$'s we use an equiangle averaging scheme. Namely we consider the set:

$$E^{\epsilon}(\mathbf{y}) = \left\{\mathbf{y}^l = \mathbf{y} + (6l \cdot \epsilon)\,\mathbf{1} \colon 0 \leq l < 1/\sqrt{\epsilon}\right\}$$

where $\mathbf{1}$ denotes the all-ones vector. Notice that since $g$ is $\lambda$-Lipschitz for constant $\lambda$, $g\left(\mathbf{y}^l\right)$ will be approximately the same for all $\mathbf{y}^l \in E^{\epsilon}(\mathbf{y})$.

**Fact 2.** *If any $\mathbf{y}^l \in E^{\epsilon}(\mathbf{y})$ is not a corner point, then at most one $\mathbf{y}^{l'} \in E^{\epsilon}(\mathbf{y})$ is a boundary point.*

*Proof.* For each dimension, at most one element in $E^{\epsilon}(\mathbf{y})$ can be $\epsilon$-close to the $(1/2)$ facet. Thus if two elements in $E^{\epsilon}(\mathbf{y})$ are boundary points, it must be because of distinct dimensions - and therefore every $\mathbf{y}^l$ is a corner point. □

Given input $\mathbf{y}$, we compute the displacement $g(\cdot)$ separately and in parallel for each $\mathbf{y}^l \in E^\epsilon$, and average at the end. Since at most one $\mathbf{y}^l$ is a boundary point, this will incur an error of at most $\sqrt{\epsilon}$.

In the generalized circuit we can construct $E^\epsilon$ using $(1/\sqrt{\epsilon})$ auxiliary nodes and $G_\zeta$ and $G_+$ gates:

$$\mathbf{x}\left[y_i^l\right] = \min\left\{\mathbf{x}\left[y_i^0\right] + (6l \cdot \epsilon), 1\right\} \pm 2\epsilon$$

## 5.3 Computing the displacement

For each $\mathbf{y}^l \in E^\epsilon$, we construct a disjoint circuit that approximates the displacement $g\left(\mathbf{y}^l\right)$. In the description of the circuit below we omit the index $l$.

**Lemma 2.** *The circuit below $O(\sqrt{\epsilon})$-approximately simulates the computation of the Hirsch et al displacement:*

1. *Whenever $(\mathbf{x}[y_i])_{i\in[2n+2]}$ is an interior point,*

$$\mathbf{x}\left[g_i^+\right] - \mathbf{x}\left[g_i^-\right] = g_i(\mathbf{x}[\mathbf{y}]) \pm O\left(\epsilon^{1/4}\right)$$

2. *Furthermore, whenever $(\mathbf{x}[y_i])_{i\in[2n+2]}$ is a corner point,*

$$\mathbf{x}\left[g_{2n+2}^+\right] - \mathbf{x}\left[g_{2n+2}^-\right] = \delta \pm O\left(\sqrt{\epsilon}\right)$$

   *and $\forall i < 2n+2$:*

$$\mathbf{x}\left[g_i^+\right] - \mathbf{x}\left[g_i^-\right] = 0 \pm O\left(\sqrt{\epsilon}\right)$$

*Proof.* We construct the circuit in five stages: (1) given $\mathbf{y}$, we extract $\mathbf{b}$, that is the binary representation of the corresponding subcube in $\{0,1\}^{2n+2}$; (2) we then compute whether $\mathbf{b}$ belongs to a path in $H$, and if so which are the previous and next vertices; (3) we compute the centers of the coordinate systems corresponding to the entrance and exit facets, and label them $\mathbf{z}^{in}$ and $\mathbf{z}^{out}$; (4) we project $\mathbf{y}$ to each facet, and transform this projection to the local polar coordinate systems - $(r^{in}, \mathbf{p}^{in})$; and (5) finally, we use all the information above to compute the displacement $\mathbf{g} = g(\mathbf{y})$.

The correctness of Lemma 2 follows from Claims 6-12.

**Extract $\mathbf{b} \in \{0,1\}^{2n+2}$**

Our first step is to extract the binary vector $\mathbf{b}$ which represents the subcube to which $\mathbf{y}$ belongs. In other words we want $b_i$ to be the indicator of $y_i < 1/2$. We do that by adding the following gadgets: $G_\zeta\left(1/2 \mid\mid c_{1/2}\right)$ and, for each $i$, $G_<\left(\mid y_i, c_{1/2} \mid b_i\right)$. Observe that now

$$\mathbf{x}[b_i] = \begin{cases} 0 \pm \epsilon & \mathbf{x}[y_i] < \mathbf{x}\left[c_{1/2}\right] - \epsilon \\ 1 \pm \epsilon & \mathbf{x}[y_i] > \mathbf{x}\left[c_{1/2}\right] + \epsilon \end{cases}$$

*Claim* 6. If $\mathbf{x}[\mathbf{y}]$ is an interior point, $\mathbf{x}[\mathbf{b}]$ is the correct representation (up to $\epsilon$ error) of the corresponding bits in $\{0,1\}^{2n+2}$.

**Neighbors in $H$**

Given $\mathbf{x}[\mathbf{b}]$ we can construct, using $G_\wedge$'s and $G_\neg$'s and a polynomial number of unused nodes, the circuits $S^H$ and $P^H$ that give the next and previous vertex visited by our collection of paths, $H$. The output of each circuit is represented by $2n+2$ unused nodes $\left\{P_i^H(\mathbf{b})\right\}$ and $\left\{S_i^H(\mathbf{b})\right\}$.

Recall that $H$ is defined in $\{0,1\}^{2n+1}$, so the last input bit is simply ignored (inside the picture it is always 0); the last output bit is used as follows. Our convention is that starting points and end points correspond to $P^H(\mathbf{b}) = \mathbf{b}$ and $S^H(\mathbf{b}) = \mathbf{b}$, respectively, and likewise for points that do not belong to any path. An exception to this is the $\mathbf{0}$ starting point, which will correspond to $P^H(\mathbf{0}) = (0^{2n+1}; 1)$: This is in accordance with the Hirsch et al construction, where the home subcube is constructed as if it continues a path from the subcube above it.

*Claim* 7. If $\mathbf{x}[\mathbf{b}]$ is an $\epsilon$-approximate binary vector, i.e. $\mathbf{x}[\mathbf{b}] \in ([0,\epsilon] \cup [1-\epsilon, 1])^{2n+2}$, then $\mathbf{x}\left[P^H(\mathbf{b})\right]$ and $\mathbf{x}\left[S^H(\mathbf{b})\right]$ correctly represent (up to $\epsilon$ error) the previous vertex and next vertex in $H$.

**Entrance and exit facets**

Let $b_i^{+in} = b_i \wedge \neg P_i^H(\mathbf{b})$, i.e. $b_i^{+in}$ is 1 if the path enters the current subcube via the positive $i$-th direction; define $b_i^{-in}$ analogously. Let $b_i^{in}$ denote the OR of $b_i^{+in}$ and $b_i^{-in}$.

The center of the entrance facet is constructed via $G_\zeta$, $G_{\times\zeta}$, $G_+$, and $G_-$ according to the formula:

$$z_i^{in} = \begin{cases} 1/2 - h/2 & b_i^{in} = 0 \text{ AND } b_i = 0 \\ 1/2 + h/2 & b_i^{in} = 0 \text{ AND } b_i = 1 \\ 1/2 & b_i^{in} = 1 \end{cases}$$

Construct $\mathbf{z}^{out}$ analogously.

Notice that if we know on which coordinate the path enters, in $\{0,1\}^{2n+2}$ it has only one possible direction; in the Hirsch et al hypercube this corresponds to always entering from the center (i.e. from the $y_i = 1/2$ facet). Also, if $\mathbf{b}$ corresponds to a non-trivial starting point $\mathbf{b}^{in} = \mathbf{0}$ and $\mathbf{z}^{in}$ is simply the center of the subcube (and similarly for $\mathbf{b}^{out}, \mathbf{z}^{out}$ when $\mathbf{b}$ is an end point).

*Claim* 8. If $\mathbf{x}[\mathbf{b}]$, $\mathbf{x}\left[P^H(\mathbf{b})\right]$, and $\mathbf{x}\left[S^H(\mathbf{b})\right]$ are $\epsilon$-approximate binary vectors, then $\mathbf{x}[\mathbf{z}^{in}]$ and $\mathbf{x}[\mathbf{z}^{out}]$ are $O(\epsilon)$-approximations to the centers of the entrance facet and exit facets, respectively.

### Max-norm polar coordinates

We are now ready to compute the local max-norm polar coordinates of the projections of $\mathbf{y}$ on the entrance and exit facets.

The max-norm radius is given by

$$r^{in} = \max_{i:\, b_i^{in}=0} \left| z_i^{in} - y_i \right|$$

Finding the maximum of a (length $2n+1$) vector requires some care when on each gate we can incur a constant error, the details are described in Algorithm 4.

The direction (max-norm) unit-vector, $\mathbf{p}$, is given by

$$p_i^{in} = \left(z_i^{in} - y_i\right)/r^{in}$$

Division is computed using Divide, introducing an error of $O\left(\sqrt{\epsilon}/r^{in}\right)$; this approximation suffices because for $r^{in} < h/8$, we multiply $p_i^{in}$ by $r^{in}$ when we interpolate. Also, we will use two nodes for each $p_i^{in}$ to represent the positive and negative values. We do the same for $(r^{out}, \mathbf{p}^{out})$.

*Claim* 9. If $\mathbf{x}[\mathbf{y}]$ is an interior point, then $\mathbf{x}[r^{in}]$ and $\mathbf{x}[r^{out}]$ are $O(\sqrt{\epsilon})$-approximations to the distances of $\mathbf{x}[\mathbf{y}]$ from $\mathbf{x}[\mathbf{z}^{in}]$ and $\mathbf{x}[\mathbf{z}^{out}]$, respectively. Furthermore, $\mathbf{x}[\mathbf{p}^{in}]$ and $\mathbf{x}[\mathbf{p}^{out}]$ are $O(\sqrt{\epsilon}/\mathbf{x}[r^{in}])$- and $O(\sqrt{\epsilon}/\mathbf{x}[r^{out}])$- to the unit-length vectors that point from $\mathbf{x}[\mathbf{y}]$ in the directions of $\mathbf{x}[\mathbf{z}^{in}]$ and $\mathbf{x}[\mathbf{z}^{out}]$, respectively.

**The final displacement**

Given $\mathbf{p}^{in}$ and $\mathbf{b}^{in}$, we can compute $\mathbf{g}^{in}$ for the special values of $r^{in}$. Recall that

$$\mathbf{g}^{in}\left(\langle r^{in}, \mathbf{p}^{in}\rangle\right) = \begin{cases} \delta\left(\mathbf{b}^{+in} - \mathbf{b}^{-in}\right) & r^{in} = 0 \\ -\delta \mathbf{p}^{in} & r^{in} = h/8 \\ \delta\left(\mathbf{b}^{-in} - \mathbf{b}^{+in}\right) & r^{in} = h/4 \\ \delta \boldsymbol{\xi}_{2n+2} & r^{in} = h/2 \end{cases}$$

We use Algorithm 5 to interpolate for intermediate values of $r^{in}$. We also need to interpolate between $\mathbf{g}^{in}$ and $\mathbf{g}^{out}$. The ratio at which we interpolate is exactly the ratio between the distance of $\mathbf{y}$ from the entrance and exit facets. We label the positive and negative components of this last interpolation $\mathbf{g}^{(interior)+}$ and $\mathbf{g}^{(interior)-}$, respectively.

When $\mathbf{y}$ is close to both facets, the interpolation may be inaccurate; however, in this case it is a corner point. (We remark that this seems to be the only part of the proof which requires us to set the threshold for a corner point and the final error at $\Theta\left(\epsilon^{1/4}\right)$ rather than $\Theta(\sqrt{\epsilon})$; this issue may be avoidable by a more careful construction of Algorithms 3 and 5.)

*Claim* 10. If $\mathbf{x}[\mathbf{y}]$ is an interior point of an intermediate subcube in the tube, and it is not a corner point, then $\left(\mathbf{x}\left[\mathbf{g}^{(interior)+}\right] - \mathbf{x}\left[\mathbf{g}^{(interior)-}\right]\right)$ is a $O\left(\epsilon^{1/4}\right)$-approximation of the Hirsch et al displacement $g(\mathbf{x}[\mathbf{y}])$.

**Corner points** We must ensure that if $\mathbf{y}$ is a corner point, we set $\mathbf{g}^{(corner)+} = \delta\boldsymbol{\xi}_{2n+2}$ and $\mathbf{g}^{(corner)-} = \mathbf{0}$: check over all pairs of coordinates whether $|\mathbf{x}[y_i] - 1/2| < 2\epsilon^{1/4}$ and $|\mathbf{x}[y_j] - 1/2| < 2\epsilon^{1/4}$. Let $z$ be the variable representing the OR of those $\binom{2n+2}{2}$ indicators. Interpolate (e.g. using Algorithm 5) between the $\left(\mathbf{g}^{(interior)+}, \mathbf{g}^{(interior)-}\right)$ we computed earlier and $\delta\boldsymbol{\xi}_{2n+2}$ with weights $z$ and $\neg z$. Label the result of the interpolation $\left(\mathbf{g}^{(tube)+}, \mathbf{g}^{(tube)-}\right)$.

We remark that whenever $z$ is ambiguous, i.e. the second smallest $|\mathbf{x}[y_i] - 1/2|$ is very close to $2\epsilon^{1/4}$, then we cannot predict the value of $\mathbf{x}[z]$; it can take any value in $[0,1]$. Nevertheless, in this case $\mathbf{x}[\mathbf{y}]$ is not a corner point, thus for most $\mathbf{y}^l \in E^\epsilon$, $\mathbf{x}\left[\mathbf{y}^l\right]$ is an interior point. This means that by Claim 10, we would compute the (approximately) correct interior displacement $\left(\mathbf{x}\left[\mathbf{g}^{(interior)+}\right] - \mathbf{x}\left[\mathbf{g}^{(interior)-}\right]\right)$. Since $\mathbf{x}[\mathbf{y}]$ is close to a corner point, this value is very close to $\delta\boldsymbol{\xi}_{2n+2} = \left(\mathbf{x}\left[\mathbf{g}^{(corner)+}\right] - \mathbf{x}\left[\mathbf{g}^{(corner)-}\right]\right)$. Therefore, although we don't know the value of $\mathbf{x}[z]$, we use it to interpolate between two (approximately) equal vectors - so the result is guaranteed to be (approximately) correct regardless of the value of $\mathbf{x}[z]$.

*Claim* 11. If $\mathbf{x}[\mathbf{y}]$ is a corner point, then $\left(\mathbf{x}\left[\mathbf{g}^{(tube)+}\right] - \mathbf{x}\left[\mathbf{g}^{(tube)-}\right]\right)$ is a $O(\sqrt{\epsilon})$-approximation of $\delta\boldsymbol{\xi}_{2n+2}$, and thus also a $O\left(\epsilon^{1/4}\right)$-approximation of the Hirsch et al displacement $g(\mathbf{x}[\mathbf{y}])$. Furthermore, Claim 10 continues to hold for $\left(\mathbf{g}^{(tube)+}, \mathbf{g}^{(tube)-}\right)$.

**Start/end subcubes and subcubes outside the tube** For start/end subcubes (except the home subcube) we use a slightly different (Cartesian) interpolation that yields a fixed point in the center of the subcube, and a displacement of $\delta\boldsymbol{\xi}_{2n+2}$ on all facets but the exit/entrance facet, respectively. For subcubes in the picture but outside the tube, we again set $\mathbf{g} = \delta\boldsymbol{\xi}_{2n+2}$.

Notice that we can infer the type of subcube from the following two-bits vector:

$$T = \left(\vee_i b_i^{in}, \vee_i b_i^{out}\right)$$

If $T = (0,0)$, the subcube is outside the tube; when $T = (0,1)$, we are at a start subcube, while $T = (1,0)$ corresponds to an end subcube; $T = (1,1)$ is an intermediate subcube in the tube. Finally, interpolate between the displacement for each type of subcube using $T$ and $\neg T$; label the result of the interpolation $(\mathbf{g}^+, \mathbf{g}^-)$.

*Claim* 12. If $\mathbf{x}[\mathbf{y}]$ is either an interior point or a corner point, of any subcube in the slice, then $(\mathbf{x}[\mathbf{g}^+] - \mathbf{x}[\mathbf{g}^-])$ is an $O\left(\epsilon^{1/4}\right)$-approximation of the Hirsch et al displacement $g(\mathbf{x}[\mathbf{y}])$.

□

### 5.4 Summing the displacement vectors

We are now ready to average over the displacement vectors we computed for each $\mathbf{y}^l$. Using $G_{\times\zeta}$ and $G_+$ we have that

$$\mathbf{x}\left[g_i^+\right] = \sum_{l=1}^{1/\sqrt{\epsilon}} \left(\sqrt{\epsilon}\mathbf{x}\left[g_i^{l+}\right]\right) \pm O\left(\sqrt{\epsilon}\right) \quad \text{and} \quad \mathbf{x}\left[g_i^-\right] = \sum_{l=1}^{1/\sqrt{\epsilon}} \left(\sqrt{\epsilon}\mathbf{x}\left[g_i^{l-}\right]\right) \pm O\left(\sqrt{\epsilon}\right)$$

**Lemma 3.** *For every input* $\mathbf{x}[\mathbf{y}]$ *and every* $i \in [2n+2]$,

$$\mathbf{x}\left[g_i^+\right] - \mathbf{x}\left[g_i^-\right] = g_i\left(\mathbf{x}[\mathbf{y}]\right) \pm O\left(\epsilon^{1/4}\right)$$

*Proof.* By Fact 2, every $\mathbf{y}^l \in E^\epsilon$, except at most one, is either an interior point or a corner point. By Lemma 2, for all those $\mathbf{y}^l$, $\left(\mathbf{x}\left[g_i^{+l}\right] - \mathbf{x}\left[g_i^{-l}\right]\right)$ is at most $O\left(\epsilon^{1/4}\right)$-far from the right displacement. The single point which is neither an interior point nor a corner point increases the error by at most $O\left(\sqrt{\epsilon}\right)$, as does the summation above. Finally, because $g$ is $\lambda$-Lipschitz for constant $\lambda$, the error added due to the distance between the $\mathbf{y}^l$'s is again at most $O\left(\sqrt{\epsilon}\right)$. □

### 5.5 Closing the loop

Finally, for each $i \in [2n+2]$: $G_+\left(\mid y_i^1, g_i^+ \mid y_i'\right)$, $G_+\left(\mid y_i', g_i^- \mid y_i''\right)$ and $G_=\left(\mid y_i'' \mid y_i^1\right)$.

## 6 $\epsilon$-GCIRCUIT with fan-out 2

In the previous section, we proved that $\epsilon$-GCIRCUIT is PPAD-complete for some constant $\epsilon > 0$. Each generalized circuit gate has fan-in at most 2, which would eventually correspond to a bound on the *incoming* degree of each player in the graphical game. In order to bound the total degree (as well as for the application to A-CEEI), we need to also bound fan-out of each gate.

**Theorem 2** (Generalized circuit)**.** *There exists a constant* $\epsilon > 0$ *such that* $\epsilon$-GCIRCUIT *with fan-out* 2 *is* PPAD*-complete.*

*Proof.* We present a black-box reduction that converts a general $\epsilon'$-GCIRCUIT instance to an instance of $\epsilon$-GCIRCUIT with fan-out 2, for $\epsilon' = \Theta\left(\sqrt{\epsilon}\right)$. Daskalakis et al [DGP09] bound the fan-out of the generalized circuit by introducing a binary tree of $G_=$ gates. Unfortunately, this blows up the error: each $G_=$ gate

**Algorithm 6** REAL2UNARY$\left(\mid a \mid b_1, \ldots b_{4/\epsilon'}\right)$

1. $G_=\left(\mid a \mid c_0\right)$
2. **for** each $k \in [4/\epsilon']$:
   (a) $G_=\left(\mid c_{k-1} \mid c_k\right)$
   # The $c_k$'s are simply copies of $a$, ensuring that each gate has fan-out at most 2:
   $$\forall k \;\; \mathbf{x}\left[c_k\right] = \mathbf{x}\left[a\right] \pm \left(k+1\right)\epsilon$$
   (b) $G_\zeta\left(k\epsilon'/4 \mid\mid \zeta_k\right)$
   (c) $G_<\left(\mid \zeta_k, c_k \mid b_k\right)$
   # The vector $(b_k)$ is the unary representation of $a$:
   $$\forall i \;\; \left(\max_{k\colon \mathrm{X}[b_k]>\epsilon} k\epsilon'/4\right) = \mathbf{x}\left[a\right] \pm \epsilon'/2$$

introduces an additive error of $\epsilon$, so the increase is proportional to the depth of the tree, i.e. $\Theta\left(\epsilon \cdot \log n\right)$. While this was acceptable for [DGP09] who used an exponentially small $\epsilon$, it clearly fails in our setting.

We overcome this obstacle by resorting to logical gates ($G_<$, $G_\vee$, $G_\wedge$, and in particular $G_\neg$). Recall that the logical gates are at most $\epsilon'$-far from the correct Boolean value. Therefore, concatenating multiple logical gates does not amplify the error. In particular, if any logical gate has a large fan-out, we can distribute its output using a binary tree of $G_\neg$ gates (we use trees of even depth). When the gate is arithmetic ($G_\zeta$, $G_{\times\zeta}$, $G_=$, $G_+$, or $G_-$), we convert its output to unary representation over $\Theta\left(\epsilon'\right)$ logical gates (Algorithm 6). Then, we copy the unary representation using trees of $G_\neg$ gates. Finally, we use $G_{\times\zeta}$ and $G_+$ gates to convert each copy of the unary representation back to a real value in $[0,1]$ (Algorithm 7).

It is interesting to note that for constant $\epsilon$ and $\epsilon'$, the unary representation has constant size, so the number of new gates is proportional to the original fan-out (i.e. the number of leaves of the binary tree that copies the unary representation). In particular this reduction increases the size of the circuit by a factor of $\Theta\left(1/\epsilon'\right)$.

**Algorithm 7** UNARY2REAL$\left(\mid b_1, \ldots b_{4/\epsilon'} \mid a'\right)$

1. $G_\zeta\left(0 \mid\mid d_0\right)$
2. **for** each $k \in [4/\epsilon']$:
   (a) $G_{\times\zeta}\left(\epsilon'/4 \mid b_k \mid c_k\right)$
   # The sum of the $c_k$'s is approximately $a$:
   $$\sum \mathbf{x}\left[c_k\right] = \mathbf{x}\left[a\right] \pm \left(\epsilon'/2 + \epsilon \cdot 4/\epsilon'\right)$$
   (b) $G_+\left(\mid d_{k-1}, c_k \mid d_k\right)$
3. $G_=\left(\mid d_k \mid a'\right)$
   # $a'$ approximately recovers the original $a$:
   $$\mathbf{x}\left[a'\right] = \mathbf{x}\left[a\right] \pm \epsilon'$$

□

# 7 From $\epsilon$-GCIRCUIT to $\epsilon^2$-ANE

The last step in our proof is to reduce the problem of finding an $\epsilon$-approximate solution to the generalized circuit to that of finding an $\Theta\left(\epsilon^2\right)$-ANE. First, we reduce to the problem of finding an $\epsilon$-WSNE in a (polymatrix, degree 3) graphical game. This reduction is implicit in Daskalakis et al [DGP09][5].

**Definition 6.** Given a polymatrix game over a degree $d$ graph with $k$ actions for each player, $\epsilon$-WSNE-POLYMATRIX$(d, k)$ is the problem of finding an $\epsilon$-Well Supported Nash Equilibrium.

**Theorem.** *(Essentially [DGP09])* $2\epsilon$-GCIRCUIT *with fan-out* $2 \leq_P \epsilon$-WSNE-POLYMATRIX$(3, 2)$

Daskalakis et al construct a different gadget for each type of gate. For example, we describe their gadget for the $G_{\times\zeta}$ gate below.

[5] In [DGP09], polymatrix games are called *games with additive utility functions*.

**Lemma 4.** *($G_{\times\zeta}$ gadget, [DGP09])*

*Let $v_1$,$v_2$, and $w$ be players in a graphical game, and suppose that the payoffs of $v_2$ and $w$ are as follows.*

**Payoff for $v_2$:**

| | *w plays* 0 | *w plays* 1 |
|---|---|---|
| $v_2$ *plays* 0 | 0 | 1 |
| $v_2$ *plays* 1 | 1 | 0 |

**Payoffs for $w$:**

**game with $v_1$:**

| | $v_1$ *plays* 0 | $v_1$*plays* 1 |
|---|---|---|
| *w plays* 0 | 0 | $\zeta$ |
| *w plays* 1 | 0 | 0 |

**game with $v_2$:**

| | $v_2$ *plays* 0 | $v_2$*plays* 1 |
|---|---|---|
| *w plays* 0 | 0 | 0 |
| *w plays* 1 | 0 | 1 |

*Then, in every $\epsilon$-WSNE $\mathbf{p}[v_2] = \min(\zeta\mathbf{p}[v_1], 1) \pm \epsilon$, where $\mathbf{p}[u]$ denotes the probability that $u$ assigns to strategy 1.*

*Proof.* (Sketch) If $\mathbf{p}[v_2] > \zeta\mathbf{p}[v_1] + \epsilon$, then in every $\epsilon$-WSNE $\mathbf{p}[w] = 1$, which contradicts $\mathbf{p}[v_2] > \epsilon$. Similarly, if $\mathbf{p}[v_2] < \min(\zeta\mathbf{p}[v_1], 1) - \epsilon$, then $\mathbf{p}[w] = 0$, which yields a contradiction when $\mathbf{p}[v_2] < 1 - \epsilon$. □

## 7.1 From $\sqrt{\epsilon}$-WSNE to $\Theta(\epsilon)$-ANE

The reduction above shows hardness for the slightly stronger notion (therefore weaker hardness) of $\epsilon$-WSNE. Daskalakis et al [DGP09] show a reduction from $\sqrt{\epsilon}$-WSNE to $\Theta(\epsilon)$-ANE for games with a constant number of players. It is easy to see that the same reduction holds for graphical games with constant degree. We sketch the proof below.

**Lemma 5.** *(Essentially [DGP09]) Given an $\epsilon$-ANE of a graphical game with payoffs in $[0,1]$ and incoming degree $d_{in}$, we can construct in polynomial time a $\sqrt{\epsilon} \cdot (\sqrt{\epsilon} + 1 + 4d_{in})$-WSNE.*

*Proof.* Let $V$ be the set of players, where each $v \in V$ has utility $U^v$ and action set $A^v$. Let $\mathbf{x} = (x_a^v) \in \Delta(\times_v A^v)$ be an $\epsilon$-ANE. Let $U_a^v(\mathbf{x}^{-v})$ denote the expected payoff for playing $a$ when all other players play according to $\mathbf{x}$. Note that $U_a^v\left(\mathbf{x}^{-v}\right) = U_a^v\left(\mathbf{x}^{\mathcal{N}_{in}(v)}\right)$ depends only on the distributions of the players in the incoming neighborhood of $v$, which we denote $\mathcal{N}_{in}\left(v\right)$. Finally, let $U_{\max}^v\left(\mathbf{x}^{-v}\right) = \max_{a\in A^v} U_a^v\left(\mathbf{x}^{-v}\right)$.

Let $k = k\left(\epsilon\right) > 0$ be some large number do be specified later. We construct our new approximate equilibrium by taking, for each player only the strategies that are within $\epsilon k$ of the optimum:

$$\hat{x}_a^v = \begin{cases} \frac{x_a^v}{1-z^v} & U_a^v\left(\mathbf{x}^{-v}\right) \geq U_{\max}^v\left(\mathbf{x}^{-v}\right) - \epsilon k \\ 0 & \text{otherwise} \end{cases}$$

where $z^v$ is the total probability that player $v$ assigns to strategies that are more than $\epsilon k$ away from the optimum.

The above division is well-defined because for $k > 1$, $z^v$ is bounded away from 1. Moreover, the following claim from [DGP09] formalizes the intuition that when $k$ is sufficiently large, the total weight on actions removed is small, so $\hat{\mathbf{x}}^v$ is close to $\mathbf{x}^v$:

*Claim* 13. (Claim 6 of [DGP09])

$$\forall v \in V \quad \sum_{a\in A^v} \left|\hat{x}_a^v - x_a^v\right| \leq \frac{2}{k-1}$$

Now, the total change to the expected payoff to player $v$ for each action $a$, is bounded by the total change in mixed strategies of its *incoming neighbors*:

$$\begin{aligned} \left|U_a^v\left(\mathbf{x}^{-v}\right) - U_a^v\left(\hat{\mathbf{x}}^{-v}\right)\right| &= \left|U_a^v\left(\mathbf{x}^{\mathcal{N}_{in}(v)}\right) - U_a^v\left(\hat{\mathbf{x}}^{\mathcal{N}_{in}(v)}\right)\right| \\ &\leq \sum_{\mathbf{a}\in A^{\mathcal{N}(v)}} \left|\hat{x}_{\mathbf{a}}^{\mathcal{N}_{in}(v)} - x_{\mathbf{a}}^{\mathcal{N}_{in}(v)}\right| \leq \sum_{w\in\mathcal{N}(v)} \sum_{a\in A^w} \left|\hat{x}_a^w - x_a^w\right| \leq \frac{2d_{in}}{k-1} \end{aligned}$$

It follows that $\hat{x}_a^v$ is a $\left(k\epsilon + \frac{4d_{in}}{k-1}\right)$-WSNE:

$$U_a^v\left(\hat{\mathbf{x}}^{-v}\right) \geq U_a^v\left(\mathbf{x}^{-v}\right) - \frac{2d_{in}}{k-1} \geq U_{\max}^v\left(\mathbf{x}^{-v}\right) - \epsilon k - \frac{2d_{in}}{k-1} \geq U_{\max}^v\left(\hat{\mathbf{x}}^{-v}\right) - \epsilon k - \frac{4d_{in}}{k-1}$$

Finally, take $k = 1 + 1/\sqrt{\epsilon}$ to get that

$$k\epsilon + \frac{4d_{in}}{k-1} \leq \sqrt{\epsilon}\cdot\left(\sqrt{\epsilon} + 1 + 4d_{in}\right)$$

□

# 8 $\epsilon$-approximate Bayesian Nash equilibrium

In this section we formally introduce and prove our corollary for $\epsilon$-approximate Bayesian Nash equilibrium in two-player games with incomplete information.

In a game with incomplete information, each player $i$ has a type $t_i$ known only to her, and the players' types $\mathbf{t} = (t_1, t_2)$ are drawn from a joint distribution which is known to everyone. The payoff for player $i$ is a function $u_i(\mathbf{a}, t_i)$ of her own type and all the players' actions.

**Definition 7.** (e.g. [SSW04])

In a *Bayesian Nash equilibrium*, for every player $i$ and every type $t_i$, the mixed strategy $x_i(t_i)$ must be a best response in expectation over the other players' types and actions:

$$\mathbb{E}_{\mathbf{t}|t_i}\left[\mathbb{E}_{\mathbf{a}\sim(x_i(t_i),x_{-i}(t_{-i}))}\left[u_i(\mathbf{a};t_i)\right]\right] \geq \max_{x_i'(t_i)\in\Delta A_i} \mathbb{E}_{\mathbf{t}|t_i}\left[\mathbb{E}_{\mathbf{a}\sim\left(x_i'(t_i),x_{-i}(t_{-i})\right)}\left[u_i(\mathbf{a};t_i)\right]\right].$$

Similarly, in an *$\epsilon$-approximate Bayesian Nash equilibrium*, for every player $i$ and every type $t_i$, the mixed strategy $x_i(t_i)$ must be an $\epsilon$-best response in expectation over the other players' types and actions:

$$\mathbb{E}_{\mathbf{t}|t_i}\left[\mathbb{E}_{\mathbf{a}\sim(x_i(t_i),x_{-i}(t_{-i}))}\left[u_i(\mathbf{a};t_i)\right]\right] \geq \max_{x_i'(t_i)\in\Delta A_i} \mathbb{E}_{\mathbf{t}|t_i}\left[\mathbb{E}_{\mathbf{a}\sim\left(x_i'(t_i),x_{-i}(t_{-i})\right)}\left[u_i(\mathbf{a};t_i)\right]\right]-\epsilon.$$

Before we prove our main corollary for incomplete information games, it is helpful to prove the following slightly weaker statement, for two players with many strategies.

**Lemma 6.** *There exists a constant $\epsilon > 0$, such that given a two-player game with incomplete information where each player has $n$ actions, finding an $\epsilon$-approximate Bayesian Nash equilibrium is* PPAD*-complete.*

*Proof.* We reduce from a bipartite polymatrix game, and let the typeset for each of the two players in the incomplete information game correspond to one side of the bipartite graph. The utility of player $i$ on edge $(i, j)$ of the polymatrix game depends on her identity $(i)$, as well as the identity $(j)$ of the vertex on the other side of that edge. We use the types of the incomplete information game, to encode $i$. We encode $j$ using the strategies of the second player in the incomplete information game.

In more detail, consider a bipartite polymatrix game for which it is PPAD-hard to compute a $4\epsilon$-approximate Nash equilibrium. Use an affine transformation to change all the payoffs from $[0,1]$ to $[1/2,1]$. It is PPAD-hard to find a $2\epsilon$-approximate Nash equilibrium in the transformed game.

We now construct the two-player incomplete information game: As we hinted before, we let the typeset of each player correspond to the vertices on one side of the bipartite graphical game. Player $i$ has $|T_i|$ types and $2\,|T_i|$ strategies, where each strategy corresponds to a pair of vertex and strategy for that vertex. If a player plays a strategy whose vertex does not match her type, her payoff is 0. Therefore in every $\epsilon$-approximate Bayesian Nash equilibrium, every player, on every type, plays the two strategies that correspond to her type with probability at least $1-2\epsilon$.

Let the joint distribution over types be as follows: pick a random edge in the bipartite graph, and assign the types corresponding to its vertices. Whenever both players play strategies that match their respective types, their payoffs are the payoffs in the (transformed) bimatrix game on that edge. In every $\epsilon$-approximate Bayesian Nash equilibrium, every player, on every type, plays a mixed strategy which is $\epsilon$-best response. Since the other player plays strategies that correspond to the correct vertex with probability at least $1-2\epsilon$, the same mixture must be a $2\epsilon$-best response for the vertex player in the bipartite game. □

In order to prove the main corollary, we need to reduce the number of actions in the above construction. Observe that we don't need each player to choose an action that uniquely identifies her type. Rather, it suffices to specify which neighbor of the other player's vertex is chosen. This can be done concisely via a coloring of the vertices such that every pair of vertices of distance exactly two have different colors; i.e. a coloring of the square of the polymatrix game's graph. The squared graph has degree $3 \cdot (3-1) = 6$, and therefore we can efficiently find a 7-coloring. It suffices for each player to specify one of 7 colors, together with one of 2 strategies for the vertex with that color. Therefore we can encode this choice using only 14 strategies.

**Corollary 1.** *There exists a constant $\epsilon > 0$, such that given a two-player game with incomplete information where each player has 14 actions, finding an $\epsilon$-approximate Bayesian Nash equilibrium is* PPAD*-complete.*

## 9 Relative $\epsilon$-Well Supported Nash equilibrium

In this section we formally introduce and prove our result for relative $\epsilon$-Well Supported Nash Equilibrium in two-player games.

As we mentioned in Section 3.1, we say that a vector of mixed strategies $\mathbf{x} \in \times_j \Delta A_j$ is a Nash equilibrium if every strategy $a_i$ in the support of $x_i$ is a best response to the actions of the mixed strategies of the rest of the players, $x_{-i}$;

$$\mathbb{E}_{a_{-i} \sim x_{-i}} \left[ u_i \left( a_i, a_{-i} \right) \right] = \max_{a' \in A_i} \mathbb{E}_{a_{-i} \sim x_{-i}} \left[ u_i \left( a', a_{-i} \right) \right].$$

In this section, we study the multiplicative relaxation of this condition. Namely,

**Definition 8.** We say that $\mathbf{x}$ is a *relative $\epsilon$-Well Supported Nash Equilibrium* (*relative $\epsilon$-WSNE*) if each $a_i$ in the support of $x_i$ is a relative $\epsilon$-best response to $x_{-i}$. I.e. $\forall a_i \in \operatorname{Supp}(x_i)$,

$$\mathbb{E}_{a_{-i} \sim x_{-i}} \left[ u_i \left( a_i, a_{-i} \right) \right] \geq (1 - \epsilon) \max_{a' \in A_i} \mathbb{E}_{a_{-i} \sim x_{-i}} \left[ u_i \left( a', a_{-i} \right) \right].$$

**Corollary 2** (Relative $\epsilon$-WSNE)**.** *There exists a constant $\epsilon > 0$ such that finding a relative $\epsilon$-Well Supported Nash Equilibrium in a bimatrix game with positive payoffs is PPAD-complete.*

*Proof.* We reduce from a bipartite, degree 3, polymatrix game (Theorem 1). The payoff bimatrix is the sum of a main game that simulates the polymatrix game and an imitation game that forces the two players to randomize (approximately) uniformly across all the polymatrix game nodes.

**Main game** As in the reduction to games of incomplete information, let each player "control" the nodes on one side of the bipartite game graph. Namely, let $n$ be the number nodes on each side of the graph; each player has $2n$ actions, each corresponding to a choice of a node and an action for that node. (We assume without loss of generality that the number of nodes is equal and every node has exactly 3 neighbors.) When the players play strategies that correspond to adjacent nodes in the graphs, they receive the payoffs of the corresponding nodes, scaled by a small positive constant, e.g. $\eta = 0.01$, to fit in $[0, \eta]$. If the nodes played do not share an edge in the bipartite game graph, the utility for both players is zero.

**Imitation game** We also let the players play a hide-and-seek game with payoffs in $\{0,1\}$ over a space of $n$ actions. Fix an arbitrary ordering over the nodes on each side of the polymatrix game; action $i$ in the imitation game corresponds to the two rows (respectively, two columns) for node $i$ in the main game. The row player tries to imitate the column player; namely her payoff is 1 if they both play their respective node $i$, and 0 otherwise. The column player tries to imitate the row player +1; her payoff is 1 if the row player plays node $i$, and she plays node $i+1 \pmod n$ (we henceforth drop the $\pmod n$ for simplicity of notation).

### Structure and value of a relative $\epsilon$-WSNE

We first claim that in every relative $\epsilon$-WSNE $(x, y)$, the two players randomize approximately uniformly across all their nodes. Let $x(i)$ denote the total probability that the row player assigns to node $i$, and analogously for $y(i)$.

**Lemma 7.** *In every $(x,y)$ relative $\epsilon$-WSNE, $x(i), y(i) \in [(1-4\eta)/n, (1+5\eta)/n]$ $\forall i$.*

*Proof.* Let $t_R = \max_i x(i)$ and $i_R^* = \arg\max_i x(i)$; we first observe that if $x(j) \le (1-4\eta)\, t_R < (1-\epsilon-3\eta)\, t_R$, then $y(j+1) = 0$. The column player's payoff from the imitation game, when playing node $j+1$, is at most $x(j)$, while from the main game she can gain at most $x(k) \cdot \eta$ for each row player's node $k$ in the neighborhood of the column player's $j+1$. Since each node has only three neighbors and they all satisfy $x(k) \le t_R$, her payoff for playing either strategy of node $j+1$ is bounded by $x(j) + 3t_R \cdot \eta < (1-\epsilon)\, t_R$. However, she can guarantee a payoff of at least $t_R$ by playing node $i_R^*$; therefore by the well-supported condition, $y(j+1) = 0$. Similarly, if we let $t_C = \max_i y(i)$, then whenever $y(j) \le (1-4\eta)\, t_C$, $x(j) = 0$.

Next, we claim that both players have full support in the imitation game, i.e. $x(i), y(i) > 0$ $\forall i \in [n]$. Assume by contradiction that this is false, and let, without loss of generality, $x(i) = 0$. Since $x(i) < (1-4\eta)/n \le (1-4\eta)\, t_R$, we also have $y(i+1) = 0$. However, this implies $x(i+1) = 0$, which in turn implies $y(i+2) = 0$, and so forth. By induction we get that $x(i), y(i) = 0$ $\forall i \in [n]$, which of course cannot be true as their sum is 1.

Therefore, for every $i \in [n]$, we also have that

$$x(i) > (1-4\eta)\, t_R \ge (1-4\eta)\, \mathsf{E}_j x(j) = (1-4\eta)/n, \tag{5}$$

and similarly $y(i) > (1-4\eta)/n$.

Finally, we claim that the maximum probability assigned to any node is at most $t_R, t_C \leq 1/((1-4\eta)n) < (1+5\eta)/n$. Otherwise, $\mathsf{E}_i x(i) < (1-4\eta)t_R$, which in particular implies that for some $i \in [n]$, $x(i) < (1-4\eta)t_R$, violating (5) □

**Corollary 6.** *In every $(x, y)$ relative $\epsilon$-WSNE, the expected utilities of both players for playing any strategy is in $[(1-4\eta)/n, (1+9\eta)/n]$.*

### Completing the proof of Corollary 2

Given a relative $\epsilon$-WSNE $(x, y)$, we can take, for each node $i$, the mixed strategy induced by the probabilities $x(i:1)/x(i)$ and $x(i:2)/x(i)$ that the row player assigns to each action (respectively, $y(j:1)/y(j)$ and $y(j:2)/y(j)$ assigned by the column player). We claim that this strategy profile is an $\epsilon'$-approximate equilibrium for the polymatrix game for some small constant $\epsilon' = \Theta(\epsilon)$ (which is PPAD-hard to find by Theorem 1).

Assume by contradiction that this is not the case. Then without loss of generality there exists a node $i$ which is controlled by the row player, and, under the induced strategy profile, can improve her payoff by $\epsilon'$ by deviating from strategy 1 to strategy 2. Therefore in the two player game, the expected payoff to the row player from playing $x(i:2)$ is greater than the payoff from playing $x(i:1)$ by at least $\epsilon' \cdot \eta \cdot \frac{1}{n} \cdot (1 - O(\eta))$. (We multiply $\epsilon'$ by the scaling factor $\eta$ and the probability $\frac{1}{n} \cdot (1 - O(\eta))$ of observing any of its neighbors.) Yet, by Corollary 6 the total payoff from playing $x(i:2)$ is at most $(1+9\eta)/n$. Therefore, the row player improves her expected payoff by a factor of $1 + \Omega(\epsilon' \cdot \eta) \geq 1 + 2\epsilon$ when deviating from $x(i:1)$ to $x(i:2)$ - but this contradicts our assumption that $(x, y)$ is a relative $\epsilon$-WSNE. □

# 10 Non-monotone markets

In this section we formally introduce our result for non-monotone markets, discuss it, and compare it to the original result of Chen, Paparas, and Yannakakis [CPY13]. A sketch of the proof appears in Appendix A.

Intuitively, a market is monotone if increasing the price of some good, while fixing the rest of the prices, never increases the excess demand for that good. Formally, we have the following definition by Chen et al:

**Definition 9.** ([CPY13]) Let $M$ be a market with $k \geq 2$ goods. We say that $M$ is *non-monotone* at price vector $\boldsymbol{\pi}$ if there exist $c > 0$, and a good $g_1$ such that:

- the excess demand $Z_1(y_1, \ldots y_k)$ is a continuous function (rather than correspondence) over $\mathbf{y} \in B(\boldsymbol{\pi}, c)$;
- $Z_1(\boldsymbol{\pi}) > 0$;
- the partial derivative of $\partial Z_1 / \partial y_1$ exists and is continuous over $B(\boldsymbol{\pi}, c)$;
- and $\partial Z_1 / \partial y_1(\boldsymbol{\pi}) > 0$.

We say that a market $M$ is *non-monotone* if there exists such a rational price vector $\boldsymbol{\pi} \geq \mathbf{0}$, and $Z_1(\boldsymbol{\pi})$ is *moderately computable*; i.e. for any $\gamma > 0$, $Z_1(\boldsymbol{\pi})$ can be approximated to within $\gamma$ in time polynomial in $1/\gamma$.

In general we want to talk about non-monotone families of utility functions, i.e. ones that support non-monotone markets. Formally,

**Definition 10.** ([CPY13]) We say that a family $\mathcal{U}$ of utility functions is *non-monotone* if:

- $\mathcal{U}$ is countable;
- if $u \colon [0, \infty)^k \to \mathbb{R}$ is in $\mathcal{U}$, then so is $u'(x_1, \ldots x_m) = a \cdot u(x_{l_1}/b_1, \ldots x_{l_k}/b_k)$ for any indices $l_1, \ldots, l_k \in [m]$ and positive (rational) $a, b_1, \ldots, b_k$;
- $u(\boldsymbol{x}) = g^*(x_i)$ is in $\mathcal{U}$ for some strictly increasing $g \colon [0, \infty) \to \mathbb{R}$; and
- there exists a non-monotone market $M_{\mathcal{U}}$ with utilities from $\mathcal{U}$.

We need to include one more definition: that of $\epsilon$-tight market equilibrium.

**Definition 11.** A price vector $\boldsymbol{\pi}$ is an $\epsilon$-tight approximate market equilibrium of $M$ if there exists a $\mathbf{z} \in Z(\boldsymbol{\pi})$ (the excess demand at $\boldsymbol{\pi}$) such that for every good $j$, $|z_j| \leq \epsilon W_j$, where $W_j$ is the sum of the endowments of good $j$.

Our main result for non-monotone markets equilibria is now formally defined:

**Corollary 3** (Non-monotone markets)**.** *Let $\mathcal{U}$ be any non-monotone family of utility functions. There exists a constant $\epsilon_{\mathcal{U}} > 0$ such that given a market $M$ where the utility of each trader is either linear or taken from $\mathcal{U}$, finding an $\epsilon_{\mathcal{U}}$-tight approximate market equilibrium is* PPAD*-hard.*

## 10.1 Why are non-monotone markets hard?

Before delving into the details of the construction, we attempt to reach some intuition: why should we expect equilibrium computation to be hard in non-monotone markets? Probably the most intuitive algorithm for finding market equilibrium is via tatonnement: raise the prices of over-demanded goods, and decrease the prices of under-demanded goods. For many markets, the tatonnement process is also computationally efficient [CMV05]. One obvious problem is that when the market is non-monotone, the tatonnement step actually takes us further away from equilibrium. However, the non-monotonicity is only local: if we set the (relative) price of the non-monotone good high enough, even the most enthusiastic traders can only afford a small amount.

The "real" reason that tatonnement fails to converge efficiently for non-monotone markets is a little more subtle. What happens when the demand for the non-monotone good $g$ increases by a factor of $(1 + \delta)$ for some small $\delta$? The tatonnement increases the price of $g$, which further increases the demand. Eventually, the price is high enough, and the demand is reduced; but due to the non-monotonicity we may have to increase the price by larger factor, i.e. $(1 + \delta')$ for $\delta' > \delta$. Now, another trader with a positive endowment of $g$ has increased her spending budget by $(1 + \delta')$, further increasing the demand for yet another good (by a larger factor). Thus a small change in the demand for one good may cause a much larger change in the demand for another good. Exploiting this "butterfly effect" lies at the heart of Chen et al's construction.

## 10.2 High-level structure of the proof

Our reduction from polymatrix games to non-monotone markets closely follows the footsteps of [CPY13]. To gain some intuition, consider two goods $g_{2i-1}$ and $g_{2i}$ for each player $i$, corresponding to her two available strategies (soon each of those goods will become a subset of goods). Let $\pi(g_{2i-1})$ and $\pi(g_{2i})$ denote their corresponding prices; those prices correspond to the probabilities that player $i$ assigns to her respective strategies. For every $i, j \in [n]$, we add a trader who is interested in selling $g_{2i-1}$ and buying $g_{2j-1}$ (and similarly for $(2i, 2j-1)$, $(2i-1, 2j)$, and $(2i, 2j)$). This trader has an endowment of $g_{2i-1}$ that is proportional to $P_{2i-1,2j-1}$, the utility of player $j$ in the bimatrix game with player $i$, when they both play the first strategy. Qualitatively, if the price $\pi(g_{2i-1})$ is high (player $i$ assigns a high probability to her first strategy), and $P_{2i-1,2j-1}$ is high (player $j$ receives a high utility from playing her first strategy to $i$'s first strategy), then the demand for good $g_{2j-1}$ is high - implying a high price in every approximate market equilibrium (i.e. player $j$ indeed assigns a high probability to her first strategy).

In order to turn this qualitative intuition into a proof we use a more complex construction. The main difficulty comes from the need to amplify the effect of a higher income for one trader on the incomes of other traders. To this end we consider, for each $i \in [n]$, two sequences of goods: $g_{2i-1} = g_{2i-1,0}, g_{2i-1,1}, \ldots, g_{2i-1,4t} = h_{2i-1}$ and $g_{2i} = g_{2i,0}, g_{2i,1}, \ldots, g_{2i,4t} = h_{2i}$. The trader mentioned in the previous paragraph actually owns $P_{2i-1,2j-1}$ units of good $h_{2i-1}$; she is still interested in good $g_{2j-1}$. Now we construct (Lemma 14) a chain of gadgets that use copies of the non-monotone markets in $\mathcal{U}$ to amplify the small gap guaranteed between $\pi(g_{2j-1})$ and $\pi(g_{2j})$ to a larger gap between $\pi(h_{2j-1})$ and $\pi(h_{2j})$.

Additionally, we want to bound the range that these prices may take. In Lemma 11 we use a price regulating gadget [CDDT09, VY11] to control the relative prices of $\pi(g_{2i-1,j})$ and $\pi(g_{2i,j})$. In Lemma 13 we show that the sums $\pi_{i,j} = \pi(g_{2i-1,j}) + \pi(g_{2i,j})$ are approximately equal. Finally, in Section A.5 we combine these lemmata to formalize a quantitative version of the qualitative intuition described above.

## 10.3 Adaptations for constant factor inapproximability

As mentioned in the introduction, Corollary 3 has a weakness in comparison to the results of Chen et al [CPY13]: it only applies to *tight* approximate market equilibrium.

Maintaining the constant hardness of approximation through most of [CPY13]'s proof is rather straightforward, but there are a few hurdles along the way. To understand the first obstacle, we must consider a subtle issue of normalization. Chen et al normalize the bimatrix game between every pair of players to have an average value of 1/2. While this does not change the absolute utility gained from any deviation, the relative utility from deviation is now divided by a factor of $\Theta(n)$. In contrast, in Theorem 1 we prove hardness for a constant $\epsilon$ when normalizing with respect to a constant degree (3), i.e. each player participates in only a constant number of bimatrix games. We overcome this difficulty by using a different normalization: only edges (i.e. bimatrix games) that belong to the game graph will have an average utility of 1/2, while the utilities on other edges remains 0. Since we proved hardness for a degree 3 graphical game, the normalization only costs us a constant factor.

More serious complications arise when trying to prove [CPY13]'s Lemma 3[6] for a constant $\epsilon$. This lemma says that certain prices (in fact, these are sums of prices), denoted $\pi_{i,0}$ for $i \in [n]$, are approximately equal. A key step in the proof of [CPY13] is to show, roughly, that in every $\epsilon(n)$-approximate market equilibrium,

$$\pi_{i,0} \geq \frac{1}{n} \sum_{j \in [n]} \pi_{j,0} - O(\epsilon(n))$$

When $\epsilon(n)$ is polynomially small, this immediately implies that $\min_{i \in [n]} \pi_{i,0}$ is within $O(\epsilon(n))$ of the average, and therefore it must also be that $\max_{i \in [n]} \pi_{i,0}$ is within $O(n \cdot \epsilon(n))$ of the average. When taking a larger $\epsilon(n)$, this reasoning breaks. The first modification we make to overcome this obstacle, is to require $\epsilon(n)$-*tight* approximate market equilibrium. This gives a two-sided bound:

$$\left| \pi_{i,0} - \frac{1}{n} \sum_{j \in [n]} \pi_{j,0} \right| = O(\epsilon(n)) \tag{6}$$

A second issue that arises in the same inequality, is that with our new normalization, which depends on the game graph $G$, we can only prove that $\pi_{i,0}$ approximates the values of its neighbors, denoted $\mathcal{N}_G(i)$. In other words, (6) becomes

$$\left| \pi_{i,0} - \frac{1}{|\mathcal{N}_G(i)|} \sum_{j \in \mathcal{N}_G(i)} \pi_{j,0} \right| = O(\epsilon(n)) \tag{7}$$

In order to relate the value of $\pi_{i,0}$ to the corresponding values of the neighboring vertices, $\pi_{j,0}$'s, we consider $T$ consecutive applications of (7): $\pi_{i,0}$ is $O(T \cdot \epsilon)$-close to the expectation over $\pi_{j,0}$ where $j$ is taken from the distribution of a $T$-steps random walk on $G$ starting from $i$. For example, if $G$ is a constant degree expander, the random walk converges in $O(\log n)$ steps, yielding a $(1/\log n)$-inapproximability result.

[6]Compare with our Lemma 13. The reader may also want to refer to Lemma 6 in the full version of [CPY13].

### Achieving constant hardness

Finally, in order to achieve hardness for a constant $\epsilon$, we want a graph with constant mixing time - and this clearly cannot be done with a constant degree[7]. Instead, in Section A.2 we construct a normalized game whose graph has a constant mixing time, each vertex has degree $O(\sqrt{n})$, and yet approximating Nash equilibrium is hard for a constant $\epsilon$. In short, we take $n$ copies of the original $n$-player game (our new game has $n^2$ players). For any pair of players that play a (non-trivial) bimatrix game in the original game, we have a copy of the same bimatrix game between all $\binom{n}{2}$ pairs of their respective copies. We also add a trivial bimatrix game between every pair of players that belong to the same copy of the original game. In Section A.2 we argue that these newly added trivial edges are only a constant fraction of all edges in the new game graph, yet this graph has a constant mixing time.

# References

[Bab14] Yakov Babichenko. Query complexity of approximate nash equilibria. In *STOC*, pages 535–544, 2014.

[BLP15] Siddharth Barman, Katrina Ligett, and Georgios Piliouras. Approximating nash equilibria in tree polymatrix games. In *Algorithmic Game Theory - 8th International Symposium, SAGT 2015, Saarbrücken, Germany, September 28-30, 2015, Proceedings*, pages 285–296, 2015.

[BPR16] Yakov Babichenko, Christos H. Papadimitriou, and Aviad Rubinstein. Can almost everybody be almost happy? In *Proceedings of the 2016 ACM Conference on Innovations in Theoretical Computer Science, Cambridge, MA, USA, January 14-16, 2016*, pages 1–9, 2016.

[BR16] Yakov Babichenko and Aviad Rubinstein. *CoRR*, abs/1608.06580, 2016.

[7] In fact, it seems that a graph where the random walks starting from any pair of neighbors converge in constant time would suffice. We do not know whether such graphs can be constructed with constant degree.

[Bud11] Eric Budish. The combinatorial assignment problem: Approximate competitive equilibrium from equal incomes. *Journal of Political Economy*, 119(6):1061 – 1103, 2011.

[CCT15] Xi Chen, Yu Cheng, and Bo Tang. Well-supported versus approximate nash equilibria: Query complexity of large games. *CoRR*, abs/1511.00785, 2015.

[CDDT09] Xi Chen, Decheng Dai, Ye Du, and Shang-Hua Teng. Settling the Complexity of Arrow-Debreu Equilibria in Markets with Additively Separable Utilities. In *FOCS*, pages 273–282, 2009.

[CDT09] Xi Chen, Xiaotie Deng, and Shang-Hua Teng. Settling the complexity of computing two-player Nash equilibria. *J. ACM*, 56(3), 2009.

[CMV05] Bruno Codenotti, Benton McCune, and Kasturi Varadarajan. Market equilibrium via the excess demand function. In *Proceedings of the Thirty-seventh Annual ACM Symposium on Theory of Computing*, STOC '05, pages 74–83, New York, NY, USA, 2005. ACM.

[CPY13] Xi Chen, Dimitris Paparas, and Mihalis Yannakakis. The complexity of non-monotone markets. In *STOC*, pages 181–190, 2013. Preprint of full version is available at http://arxiv.org/abs/1211.4918v1.

[Das13] Constantinos Daskalakis. On the complexity of approximating a nash equilibrium. *ACM Transactions on Algorithms*, 9(3):23, 2013.

[DFSS14] Argyrios Deligkas, John Fearnley, Rahul Savani, and Paul G. Spirakis. Computing Approximate Nash Equilibria in Polymatrix Games. In *Web and Internet Economics - 10th International Conference, WINE 2014, Beijing, China, December 14-17, 2014. Proceedings*, pages 58–71, 2014.

[DGP09] Constantinos Daskalakis, Paul W. Goldberg, and Christos H. Papadimitriou. The complexity of computing a Nash equilibrium. *Commun. ACM*, 52(2):89–97, 2009.

[Fol67] D.K. Foley. Resource Allocation and the Public Sector. *Yale Economic Essays*, 7(1):45–98, 1967.

[GR14] Paul W. Goldberg and Aaron Roth. Bounds for the query complexity of approximate equilibria. In *EC*, pages 639–656, 2014.

[HdRS08] Sébastien Hémon, Michel de Rougemont, and Miklos Santha. Approximate Nash Equilibria for Multi-player Games. In *SAGT*, pages 267–278, 2008.

[HN13] Sergiu Hart and Noam Nisan. The query complexity of correlated equilibria. *CoRR*, abs/1305.4874, 2013.

[HPV89] Michael D. Hirsch, Christos H. Papadimitriou, and Stephen A. Vavasis. Exponential lower bounds for finding brouwer fix points. *J. Complexity*, 5(4):379–416, 1989.

[Kea07] Michael Kearns. *Graphical games*, chapter 7, pages 159–180. Cambridge University Press, 2007.

[LMM03] Richard J. Lipton, Evangelos Markakis, and Aranyak Mehta. Playing large games using simple strategies. In *EC*, pages 36–41, 2003.

[Max97] R. R. Maxfield. General equilibrium and the theory of directed graphs. *Journal of Mathematical Economics*, 27(1):23–51, 1997.

[Nas51] John Nash. Non-cooperative games. *The Annals of Mathematics*, 54:286–295, 1951.

[OPR14] Abraham Othman, Christos H. Papadimitriou, and Aviad Rubinstein. The complexity of fairness through equilibrium. In *EC*, pages 209–226, 2014.

[Oth14] Abraham Othman. *Course Match User Manual*, 1.3 edition, 2014.

[Pap94] Christos H. Papadimitriou. On the complexity of the parity argument and other inefficient proofs of existence. *J. Comput. Syst. Sci.*, 48(3):498–532, 1994.

[Rub14] Aviad Rubinstein. Computational complexity of approximate nash equilibrium in large games. *CoRR*, abs/1405.0524, 2014.

[Rub16] Aviad Rubinstein. Settling the complexity of computing approximate two-player nash equilibria. In *To appear in FOCS*, 2016.

[Shm12] Eran Shmaya. Brouwer Implies Nash Implies Brouwer. http://theoryclass.wordpress.com/2012/01/05/brouwer-implies-nash-implies-brouwer/, 2012.

[SSW04] Satinder P. Singh, Vishal Soni, and Michael P. Wellman. Computing approximate bayes-nash equilibria in tree-games of incomplete information. In *EC*, pages 81–90, 2004.

[SV12] Grant Schoenebeck and Salil P. Vadhan. The computational complexity of nash equilibria in concisely represented games. *TOCT*, 4(2):4, 2012.

[TV85] W. Thomson and H.R. Varian. Theories of justice based on symmetry. *Social Goals and Social Organizations: Essays in Memory of Elisha Pazner*, 1985.

[Var74] H. Varian. Equity, envy, and efficiency. *Journal of Economic Theory*, 9(1):63–91, 1974.

[VY11] Vijay V. Vazirani and Mihalis Yannakakis. Market equilibrium under separable, piecewise-linear, concave utilities. *J. ACM*, 58(3):10, 2011.

# A Non-monotone markets: proof of inapproximability

In this section we prove our main inapproximability result for non-monotone markets:

**Corollary 3** (Non-monotone markets)**.** *Let $\mathcal{U}$ be any non-monotone family of utility functions. There exists a constant $\epsilon_{\mathcal{U}} > 0$ such that given a market $M$ where the utility of each trader is either linear or taken from $\mathcal{U}$, finding an $\epsilon_{\mathcal{U}}$-tight approximate market equilibrium is* PPAD*-hard.*

## A.1 Normalized polymatrix games

We identify $n$-player, 2-strategy polymatrix graphical games with $2n \times 2n$ matrices by letting the $(i, j)$-th $2 \times 2$ block correspond to the payoff matrix of player $j$ in the bimatrix game with player $i$.

Given a game $\mathcal{G}$, let $\mathbf{P}'$ be the $2n \times 2n$ induced payoff matrix. We normalize $\mathbf{P}'$ as follows:

$$P_{2i,2j-1} = \begin{cases} 1/\left(2\Delta\right) + \left(P'_{2i,2j-1} - P'_{2i,2j}\right)/\left(2\Delta\right) & (i,j) \in E \\ 0 & \text{otherwise} \end{cases} \tag{8}$$

where $E$ is the edge set for the graphical game[8] and $\Delta$ is the maximum degree. We define $P_{2i,2j}$, $P_{2i-1,2j-1}$, $P_{2i-1,2j}$ analogously. Notice that $\mathbf{P}$ and $\mathbf{P}'$ have the same $\epsilon$-WSNE, up to the normalization by the degree $\Delta$. In particular finding an $(\epsilon/\Delta)$-WSNE in $\mathbf{P}$ continues to be PPAD-complete.

Observe that in this formulation, finding an $\epsilon$-WSNE is equivalent to finding a vector $\mathbf{x} \in [0,1]^{2n}$ s.t. $x_{2i-1} + x_{2i} = 1$ and

$$\begin{aligned} \mathbf{x}^\top \cdot \mathbf{P}_{2i-1} > \mathbf{x}^\top \cdot \mathbf{P}_{2i} + \epsilon \quad &\Longrightarrow \quad x_{2i} = 0 \\ \mathbf{x}^\top \cdot \mathbf{P}_{2i-1} < \mathbf{x}^\top \cdot \mathbf{P}_{2i} - \epsilon \quad &\Longrightarrow \quad x_{2i-1} = 0 \end{aligned}$$

## A.2 Games on graphs with a constant mixing time

Given the correspondence defined above between $n$-player games and $2n \times 2n$ matrices, we see that the structure of the game graph plays a non-trivial role in the construction. In particular, adding trivial edges between vertices, i.e. adding zero-utility bimatrix games between players, has no affect on the utility of the players, but changes the corresponding normalized matrix. For reasons that will become clear much later in the proof, we would like our game graph to have a constant mixing time.

Indeed, a trivial candidate with very fast mixing is the complete graph. However, such a blowup in the degree would dilute our inapproximability factor in the *normalized game*. Instead, we consider $n$ copies $(v^1, \ldots, v^n)$ of each player $v \in V$ in the original game. If players $u$ and $v$ play a bimatrix game $\mathcal{G}_{u,v}$ in the original game $\mathcal{G}$, then for every $i, j \in [n]$, we construct the same bimatrix game $\mathcal{G}_{u,v}$ between $u^i$ and $v^j$. Our game graph now consists of $n^2$ vertices, each with degree[9] $3n$. Finally, within each copy $V^i$, we add trivial edges between all the vertices not otherwise connected (including self-loops). Normalize this game using (8). We use $\overline{\mathcal{G}}$ to denote the new game and $\overline{G}$ for the new game graph; we henceforth let $\Delta = 4n - 3$ denote the degree. In the next two lemmata we show that this game satisfies the two properties we need: finding an $\epsilon$-WSNE of $\overline{\mathcal{G}}$ is PPAD-complete, and the mixing time of $\overline{G}$ is constant.

[8]Notice that this definition allows self-loops in the game graph.

[9]Theorem 1 promises a graphical game of degree at most 3. It is not hard to extend to a 3-regular graph game with only a constant loss in the approximation factor.

**Lemma 8.** *Given an $\epsilon$-WSNE in $\overline{\mathcal{G}}$, we can (efficiently) construct a $(4\epsilon/3)$-WSNE for $\mathcal{G}$.*

*Proof.* For each player $v$, we take the average of the mixed strategies of $v^1, \ldots, v^n$. The utility of $v$ is the same as the average of utilities of $v^1, \ldots, v^n$, and if $v$ has a $(4\epsilon/3)$-improving deviation, then at least one of the copies $v^i$ has an $\epsilon$-improving deviation. (The $(4/3)$ factor comes from the change in the degree.) □

**Lemma 9.** *Let $\pi_{v^i,T}$ be the distribution of a random walk on $\overline{G}$ after $T$ steps, starting from $v^i$, and let $\pi^*$ be the uniform distribution on the vertices of $\overline{G}$. Then*

$$\left\|\pi_{v^i,T} - \pi^*\right\|_1 \leq \left(\frac{1}{4}\right)^{T/2} + \left(\frac{3}{4}\right)^{T/2} = 2^{-O(T)}$$

*Proof.* At each step of the random walk, there is a constant probability (greater than 3/4) of walking on a non-trivial edge, which takes us to another (uniformly random) copy of the original game; thereafter the copy of the game remains uniformly random. Similarly, at each step there is a constant probability (greater than 1/4) of moving to a vertex within the same copy (again, uniformly random). Thus conditioned on having walked on a non-trivial edge, and then on an edge within the same copy, the distribution is uniform. Since all vertices have the same degree, this is also the stationary distribution, and we never leave it. □

For simplicity, in the following we redefine $n$ to be the size of $\overline{G}$ (and hence $\Delta \approx 4\sqrt{n}$).

## A.3 Construction

Let $N$ be a sufficiently large constant, and let $t = \log N$. Note that $N$ depends on the parameters of the non-monotone market in $\mathcal{U}$, but not on the size $n$ of our construction. We use the notation $O_N(\cdot)$ to denote the asymptotic behavior when $N$ goes to infinity (but arbitrarily slower than $n$). We prove that it is PPAD-hard to find an $\eta$-tight approximate market clearing equilibrium for $\eta = N^{-8}\epsilon$, where $\epsilon$ is the inapproximability factor from Lemma 8.

For each vertex $i \in [n]$ we construct a series of $4t + 1$ gadgets $\mathcal{R}_{i,j}$, for $j \in [0 : 4t]$. Each gadget is composed of:

**Main goods** $g_{2i-1,j}$ and $g_{2i,j}$ are the main goods in the reduction. They are used to encode the weights assigned to strategies $x_{2i-1}$ and $x_{2i}$, respectively.

**Non-monotone gadget** For each $j \in [4t]$, we include additional goods $s_{i,j,3}, \ldots, s_{i,j,k}$ and a non-monotone gadget

$$\mathbf{NM}\left(\mu, \gamma, g_{2i-1,j}, g_{2i,j}, s_{i,j,3}, \ldots, s_{i,j,k}\right)$$

This means that we scale the non-monotone market guaranteed to exist in $\mathcal{U}$ according to parameters $\gamma$ and $\mu$ such that when all the prices are approximately the same, the excess demand of $g_{2i-1,j}$ increases linearly with its price. Formally, we have the following lemma by Chen et al.

**Lemma 10.** *(Lemma 3.1 of [CPY13]) There exist two (not necessarily rational) positive constants $c$ and $d$ with the following property. Given any $\gamma > 0$, one can build a market $M_\gamma$ with utilities from $\mathcal{U}$ and goods $g_{2i-1,j}, g_{2i,j}, s_{i,j,3}, \ldots, s_{i,j,k}$ in polynomial time in $1/\gamma$ such that:*

*Let $f_{\gamma,\mu}(x)$ denote the excess demand function of $g_{2i-1,j}$ when the price of $g_{2i-1}$ is $1+x$, and the prices of all other $k-1$ goods are $1-x$. Then $f_{\gamma,\mu}$ is well defined over $[-c,c]$ with $|f_{\gamma,\mu}(0)| \leq \mu\gamma$ and its derivative $f'_{\gamma,\mu}(0) = d > 0$. For any $x \in [-c,c]$, $f_{\gamma,\mu}(x)$ also satisfies*

$$\left|f_{\gamma,\mu}(x) - f_{\gamma,\mu}(0) - \mu d x\right| \leq |\mu x / D|, \text{ where } D = \max\{20, 20/d\}.$$

Finally, we would like to set $\mu = \Delta/d$; in particular, this would imply that $f_{\gamma,\mu}(x) \approx \Delta x$. However, as mentioned above, $d$ may be irrational. Instead, let $d^*$ be a positive rational constant that satisfies

$$1 - 1/D \leq d^* \cdot d \leq 1$$

We set the parameters $\mu = d^*\Delta$ and $\gamma = 1/N^6$.

**Price regulating gadget** For $j \in [4t]$, we include a price regulating gadget

$$\mathbf{PR}\left(\tau, \alpha_j, g_{2i-1,j}, g_{2i,j}, s_{i,j,3}, \ldots s_{i,j,k}\right),$$

whereas for $j = 0$, we don't have goods $s_{i,0,3}, \ldots s_{i,0,k}$, and simply include

the gadget

$$\mathbf{PR}\left(\tau, \alpha_0, g_{2i-1,0}, g_{2i,0}\right).$$

The parameters are set to $\tau = N\Delta$ and $\alpha_i = 2^i/N^5$. Notice that $\alpha_0 = N^{-5}$ and $\alpha_{4t} = N^{-1} = \beta$.

This gadget ensures that in any approximate equilibrium, the price ratio $\pi\left(g_{2i-1,j}\right)/\pi\left(g_{2i,j}\right)$ is always in the range $\left[\frac{1-\alpha_j}{1+\alpha_j}, \frac{1+\alpha_j}{1-\alpha_j}\right]$. Furthermore, within each gadget $\mathcal{R}_{i,j}$, the prices of all the goods besides $g_{2i-1,j}$ are exactly equal:

$$\pi\left(g_{2i}, j\right) = \pi\left(s_{i,j,3}\right) = \cdots = \pi\left(s_{i,j,k}\right).$$

More specifically, we have two traders $T_1$ and $T_2$ with endowments $(k-1)\,\tau$ of $g_{2i-1,j}$, for $T_1$ and $\tau$ of each of the other goods for $T_2$. The utilities are defined as

$$\begin{aligned} u_1 &= (1+\alpha_j)\, x\left(g_{2i-1,j}\right) + (1-\alpha_j)\left(x\left(g_{2i,j}\right) + \sum_{l=3}^{k} x\left(s_{i,j,l}\right)\right) \\ u_2 &= (1-\alpha_j)\, x\left(g_{2i-1,j}\right) + (1+\alpha_j)\left(x\left(g_{2i,j}\right) + \sum_{l=3}^{k} x\left(s_{i,j,l}\right)\right). \end{aligned}$$

In particular, $T_1$ and $T_2$ do not trade whenever $\pi\left(g_{2i-1,j}\right)/\pi\left(g_{2i,j}\right) \in \left(\frac{1-\alpha_j}{1+\alpha_j}, \frac{1+\alpha_j}{1-\alpha_j}\right)$.

**Auxiliary goods** For $j = 0$, we also include an auxiliary good $\textsc{aux}_i$. Its eventual purpose is to disentangle the price of $g_{2i-1,0}$ and $g_{2i,0}$ from the utility that the actions of player $i$ causes to other players.

### Single-minded traders graph

We connect the groups of goods ($\mathcal{R}_{i,j}$'s) using the following single-minded traders. We use $(w, g_1\colon g_2)$ to denote a trader with endowment $w$ of good $g_1$ who only wants good $g_2$. Similarly, we use $(w, g_1, g_2\colon g_3)$ to denote a trader who has an endowment $w$ of each of $g_1$ and $g_2$, and only wants $g_3$.

1. For each $i \in [n]$ and $j \in [0 : 4t-1]$, we add two traders from $\mathcal{R}_{i,j}$ to $\mathcal{R}_{i,j+1}$: $(\Delta, g_{2i-1,j} : g_{2i-1,j+1})$ and $(\Delta, g_{2i,j} : g_{2i,j+1})$. These traders help propagate price discrepancies from $g_{i,0}$ to $g_{i,4t}$.

2. Recall that we use $g_i$ as short for $g_{i,0}$ and $h_i$ for $g_{i,4t}$. For each pair $(i,j) \in E$ we add the following four traders: $(\Delta P_{2i-1,2j-1}, h_{2i-1} : g_{2j-1})$, $(\Delta P_{2i,2j-1}, h_{2i} : g_{2j-1})$, $(\Delta P_{2i-1,2j}, h_{2i-1} : g_{2j})$, $(\Delta P_{2i,2j}, h_{2i} : g_{2j})$. Since $\mathbf{P}$ is normalized, we have

$$\Delta P_{2i-1,2j-1} + \Delta P_{2i-1,2j} = \Delta P_{2i,2j-1} + \Delta P_{2i,2j} = 1$$

These traders will enforce the approximate Nash equilibrium.

3. Connect the auxiliary goods: We let

$$r_{2j-1} = 2\Delta - \Delta \sum_{i \in \mathcal{N}(j)} \left(P_{2i-1,2j-1} + P_{2i,2j-1}\right) > 0$$

$$r_{2j} = 2\Delta - \Delta \sum_{i \in \mathcal{N}(j)} \left(P_{2i-1,2j} + P_{2i,2j}\right) > 0$$

note that $r_{2j-1} + r_{2j} = 2\Delta$.

We add the following traders: $((1-\beta)\, r_{2j-1}, \text{AUX}_j : g_{2j-1})$, $((1-\beta)\, r_{2j}, \text{AUX}_j : g_{2j})$, and $((1-\beta)\, \Delta, g_{2j-1}, g_{2j} : \text{AUX}_j)$.

Notice that the economy graph is strongly connected (because $G$ is strongly connected); therefore an equilibrium always exists [Max97]. The supplies and demands for each good are summarized in Table 1.

## A.4 Structure of a market equilibrium

We now prove some properties that every $\eta$-tight approximate equilibrium $\boldsymbol{\pi}$ must satisfy. Recall that $\eta = N^{-8}\epsilon$, where $\epsilon$ is the inapproximability factor for the polymatrix game.

We begin with the application of the price regulating markets:

**Lemma 11.** *For every $i \in [n]$ and $j \in [0:4t]$,*

$$\frac{1-\alpha_j}{1+\alpha_j} \le \frac{\pi\left(g_{2i-1}, j\right)}{\pi\left(g_{2i}, j\right)} \le \frac{1+\alpha_j}{1-\alpha_j}$$

*and*

$$\pi\left(g_{2i}, j\right) = \pi\left(s_{i,j,3}\right) = \cdots = \pi\left(s_{i,j,k}\right)$$

Table 1: Goods and traders

| good / [total supply] | supplied by | demanded by |
|---|---|---|
| $g_{2i-1}, g_{2i}$<br>$\left[N\Delta\left(1+o_N\left(1\right)\right)\right]$ | **PR** trader;<br>$(\Delta, g_{2i-1} : g_{2i-1,1})$,<br>$(\Delta, g_{2i} : g_{2i,1})$;<br>$((1-\beta)\Delta, g_{2i-1}, g_{2i} : \text{AUX}_i)$ | **PR** traders;<br>$(\Delta P_{2j-1,2i-1}, h_{2j-1} : g_{2i-1})$,<br>$\vdots$<br>$(\Delta P_{2j,2i}, h_{2j} : g_{2i})$;<br>$((1-\beta) r_{2i-1}, \text{AUX}_i : g_{2i-1})$,<br>$((1-\beta) r_{2i}, \text{AUX}_i : g_{2i})$ |
| $h_{2i-1}$<br>$\left[(k-1) N\Delta\left(1+o_N\left(1\right)\right)\right]$ | **PR** trader;<br>**NM** traders;<br>$(\Delta P_{2i-1,2j-1}, h_{2i-1} : g_{2j-1})$,<br>$(\Delta P_{2i-1,2j}, h_{2i-1} : g_{2j})$ | **PR** traders;<br>**NM** traders;<br>$(\Delta, g_{2i-1,4t-1} : h_{2i-1})$ |
| $h_{2i}$<br>$\left[N\Delta\left(1+o_N\left(1\right)\right)\right]$ | **PR** trader;<br>**NM** traders;<br>$(\Delta P_{2i,2j-1}, h_{2i} : g_{2j-1})$,<br>$(\Delta P_{2i,2j}, h_{2i} : g_{2j})$ | **PR** traders;<br>**NM** traders;<br>$(\Delta, g_{2i,4t-1} : h_{2i})$ |
| $g_{2i-1,j}$<br>$\left[(k-1) N\Delta\left(1+o_N\left(1\right)\right)\right]$ | **PR** trader;<br>**NM** traders;<br>$(\Delta, g_{2i-1,j} : g_{2i-1,j+1})$ | **PR** traders;<br>**NM** traders;<br>$(\Delta, g_{2i-1,j-1} : g_{2i-1,j})$ |
| $g_{2i,j}$<br>$\left[N\Delta\left(1+o_N\left(1\right)\right)\right]$ | **PR** trader;<br>**NM** traders;<br>$(\Delta, g_{2i,j} : g_{2i,j+1})$ | **PR** traders;<br>**NM** traders;<br>$(\Delta, g_{2i,j-1} : g_{2i,j})$ |
| $s_{i,j,l}$<br>$\left[N\Delta\left(1+o_N\left(1\right)\right)\right]$ | **PR** trader;<br>**NM** traders; | **PR** traders;<br>**NM** traders; |
| $\text{AUX}_i$<br>$\left[N\Delta\left(1+o_N\left(1\right)\right)\right]$ | $((1-\beta) r_{2i-1}, \text{AUX}_i : g_{2i-1})$,<br>$((1-\beta) r_{2i}, \text{AUX}_i : g_{2i})$ | $((1-\beta)\Delta, g_{2i-1}, g_{2i} : \text{AUX}_i)$ |

*Proof.* Follows from the construction of the price regulating markets. For more details see the proof[10] of Lemma 6 in the full version of [CPY13], or previous works that use similar gadgets [CDDT09, VY11]. □

We henceforth use $\pi_{i,j}$ to denote the sum of the $(i,j)$-th main goods: $\pi_{i,j} = \pi(g_{2i-1,j}) + \pi(g_{2i,j})$.

**Lemma 12.**

$$(1 - O_N(\eta))\, \pi_{i,0}/2 \leq \pi(\text{AUX}_i) \leq (1 + O_N(\eta))\, \pi_{i,0}/2$$

*Proof.* The total supply of $\text{AUX}_i$ is $2(1-\beta)\Delta$, yet the demand from the single-minded trader $((1-\beta)\Delta, g_{2i-1}, g_{2i} : \text{AUX}_i)$ is $(1-\beta)\Delta \frac{\pi_{i,0}}{\pi(\text{AUX}_i)}$. (For the upper bound we use the fact that $\pi$ is a *tight* approximate market equilibrium.) □

We are now ready to prove that the cost of every pair of main goods is approximately the same. Let $\delta = N^2\eta$.

**Lemma 13.** *Let* $\pi_{\max} = \max_{i,j} \pi_{i,j}$ *and* $\pi_{\min} = \min_{i,j} \pi_{i,j}$, *then*

$$\pi_{\max}/\pi_{\min} \leq 1 + O_N(\delta).$$

*Proof.* The proof of this lemma is the main obstacle which requires the tightness assumption of the market equilibrium, as well as our bound on the mixing time from Lemma 9.

Recall that by Lemma 11, the prices of all the goods in each gadget $\mathcal{R}_{i,j}$ are approximately equal. Thus, using our bound on the clearing error, we have that for each $(i,j) \in [n] \times [0:4t]$,

$$|\text{total spent on } \mathcal{R}_{i,j} - \text{total worth of } \mathcal{R}_{i,j}| \leq O_N(\eta \cdot kN\Delta)\, \pi_{i,j} = O_N(\eta \cdot N\Delta)\, \pi_{i,j} \tag{9}$$

By Walras' Law, the traders within each $\mathcal{R}_{i,j}$ (i.e. the price regulating and non-monotone gadgets) contribute the same to both quantities in (9). Similarly, by Lemma 12, the auxiliary traders contribute $O_N(\eta\Delta\pi_{i,0})$-approximately the same (for $j = 0$). Therefore the money spent on $\mathcal{R}_{i,j}$ by the single minded traders is approximately the same as the total worth of endowments in $\mathcal{R}_{i,j}$ of single minded traders:

[10] In their statement, Chen et al require an $\epsilon$-additively approximate equilibrium, and for a much smaller $\epsilon$. However their proof continues to hold with our parameters.

- For each $(i, j) \in [n] \times [4t]$, the restriction of (9) to the single-minded traders gives
$$|\Delta\pi_{i,j-1} - \Delta\pi_{i,j}| = O_N(\eta \cdot N\Delta)\,\pi_{i,j} \tag{10}$$
- Similarly, for each group $\mathcal{R}_{i,0}$, we have
$$\left|\sum_{l \in \mathcal{N}(i)} \pi_{l,4t} - \Delta\pi_{i,0}\right| = O_N(\eta \cdot N\Delta)\,\pi_{i,0} \tag{11}$$

Applying (10) inductively, we have that for any $i \in [n]$ and for any $j, l \in [0\colon 4t]$,

$$|\pi_{i,j} - \pi_{i,l}| = O_N(\eta \cdot Nt)\,\pi_{i,l}$$

Combining, with (11) we have,

$$\left|\pi_{i,0} - \frac{1}{\Delta}\sum_{l \in \mathcal{N}_{G^{\mathcal{G}}}(i)} \pi_{l,0}\right| = O(\eta \cdot Nt)\,\pi_{i,0}$$

Thus for each $i$, $\pi_{i,0}$ is $O(\eta \cdot Nt)$-approximately equal to the average of its neighbors in $\overline{G}$. Repeating this argument $T$ times, we have that $\pi_{i,0}$ is $O_N(T\eta \cdot Nt)$-approximately equal to the expectation over a $T$-step random walk in $\overline{G}$ starting from $i$. By Lemma 9, after $T = O(\log \delta)$ steps the random walk $\delta$-approximately converges to the uniform distribution, and we have

$$\begin{aligned}\left|\pi_{i,0} - \frac{1}{n}\sum_{l \in [n]} \pi_{l,0}\right| &= O_N(T\eta \cdot Nt)\,\frac{1}{n}\sum_{l \in [n]} \pi_{l,0} + \delta \max_{l \in [n]} \pi_{l,0} \\ &= O_N(\delta) \max_{l \in [n]} \pi_{l,0}\end{aligned}$$

□

Finally, we have the following lemma which describes the action of the non-monotone gadgets.

**Lemma 14.** *(Lemma 6 of [CPY13])*

$$\frac{1+\alpha_{j-1}}{\pi\left(g_{2i-1,j-1}\right)} = \frac{1-\alpha_{j-1}}{\pi\left(g_{2i,j-1}\right)} \implies \frac{1+\alpha_j}{\pi\left(g_{2i-1,j}\right)} = \frac{1-\alpha_j}{\pi\left(g_{2i,j}\right)} \textit{ and}$$
$$\frac{1-\alpha_{j-1}}{\pi\left(g_{2i-1,j-1}\right)} = \frac{1+\alpha_{j-1}}{\pi\left(g_{2i,j-1}\right)} \implies \frac{1-\alpha_j}{\pi\left(g_{2i-1,j}\right)} = \frac{1+\alpha_j}{\pi\left(g_{2i,j}\right)}$$

*Proof.* The demand for $g_{2i-1,j}$ and $g_{2i,j}$ comes from three sources: the single-minded traders, $(\Delta, g_{2i-1,j-1} : g_{2i-1,j})$ and $(\Delta, g_{2i,j-1} : g_{2i,j})$; the non-monotone gadget; and the price regulating gadget. Assume without loss of generality that the first premise holds, i.e. $\frac{1+\alpha_{j-1}}{\pi(g_{2i-1,j-1})} = \frac{1-\alpha_{j-1}}{\pi(g_{2i,j-1})}$. When the prices of $g_{2i-1,j}$ and $g_{2i,j}$ are equal, the demand from $(\Delta, g_{2i-1,j-1} : g_{2i-1,j})$ is larger since she has more income from $g_{2i-1,j-1}$. In order to account for this difference, $\pi\left(g_{2i-1,j}\right)$ must be higher - but then the demand from the traders in the non-monotone market increases. Thus we further have to increase $\pi\left(g_{2i-1,j}\right)$, until we reach the threshold of the price regulating traders: $\left(1+\alpha_j\right)/\left(1-\alpha_j\right)$.

Formally, normalize $\boldsymbol{\pi}$ such that $\pi_{i,j} = \pi\left(g_{2i-1,j}\right) + \pi\left(g_{2i,j}\right) = 2$. Thus by Lemma 13, $\pi_{i,j-1}$ is also $O_N(\delta)$-close to 2. Let $f(x)$ denote the excess demand from the traders in the non-monotone gadget when $\pi\left(g_{2i-1,j}\right) = 1+x$ and $\pi\left(g_{2i-1,j}\right) = \pi\left(s_{i,j,3}\right) = \cdots = \pi\left(s_{i,j,k}\right) = 1-x$ (recall from Lemma 11 that the latter prices are always equal to each other). By Lemma 10, we have that $|f(0)| \leq \mu\gamma$, and for all $x \in [-c, c]$:

$$|f(x) - f(0) - \mu d x| \leq |\mu x / D|.$$

Now, let $\pi\left(g_{2i-1,j-1}\right) = 1+y$; notice that by Lemma 13, $y = \alpha_{j-1} \pm O_N(\delta)$. Let $h(x,y)$ excess demand from all traders besides the two that belong to the price regulating gadget. Then,

$$h(x,y) = f(x) + \frac{\Delta(1+y)}{1+x} - \Delta = f(x) - \frac{\Delta x}{1+x} + \frac{\Delta y}{1+x}$$

For small $x$, we show that $f(x) \approx \Delta x/(1+x)$. More precisely,

$$\begin{aligned}
|f(x) - \Delta x/(1+x)| &\leq |f(x) - \mu d x| + |\mu d x - \Delta x| + \Delta\,|x - x/(1+x)| \\
&\leq |f(0)| + 2\,|\mu x/D| + 2\Delta x^2 \\
&\leq \Delta \cdot \left(d^* \cdot \gamma + 2x/20 + 2x^2\right) \\
&\leq \frac{\Delta y}{3}
\end{aligned}$$

where the first inequality follows from the triangle inequality; the second follows by application of Lemma 10 for the first difference and the definitions of $\mu$ and $d^*$ for the second; the third inequality applies the Lemma 10 again; finally the last inequality holds because for sufficiently large constant $N$, the parameters $\gamma$ and $x$ are sufficiently small.

Therefore, the excess demand must be balanced by the demand from the price regulating traders, implying that indeed $\frac{1+\alpha_j}{\pi(g_{2i-1,j})} = \frac{1-\alpha_j}{\pi(g_{2i,j})}$. □

## A.5 From market equilibrium to Nash equilibrium

To complete the proof of Corollary 3, we must construct an $\epsilon$-WSNE from any $\eta$-tight approximate market equilibrium.

For each $i \in [n]$, let $\theta_i = \left(\pi\left(h_{2i-1}\right) + \pi\left(h_{2i}\right)\right)/2$. We define

$$x_{2i-1} = \frac{\pi\left(h_{2i-1}\right)/\theta_i - (1-\beta)}{2\beta} \text{ and } x_{2i} = \frac{\pi\left(h_{2i}\right)/\theta_i - (1-\beta)}{2\beta} \tag{12}$$

Observe that $x_{2i-1} + x_{2i} = 1$.

Suppose that

$$\mathbf{x}^\top \cdot \mathbf{P}_1 \geq \mathbf{x}^\top \cdot \mathbf{P}_2 + \epsilon$$

We show that this forces $x_1 = 1$ and $x_2 = 0$; by the discussion in Section A.1 this implies that $\mathbf{x}$ is indeed an $\epsilon$-WSNE.

The following traders spend money on $g_1$:

1. For each $i \in \mathcal{N}_{\overline{G}}(1)$, there is a $(\Delta P_{i,1}, h_i; g_1)$ trader. The total money these traders spend on $g_1$ is

$$\sum \Delta P_{i,1} \cdot \pi\left(h_1\right) = \sum \Delta P_{i,1}\left(1 - \beta + 2\beta \cdot x_i\right)\theta_{\lceil i/2 \rceil}$$

2. For each $i \in \mathcal{N}_{\overline{G}}(2)$, there is a $(\Delta P_{i,2}, h_i; g_2)$ trader. The total money these traders spend on $g_2$ is

$$\sum \Delta P_{i,2} \cdot \pi\left(h_2\right) = \sum \Delta P_{i,2}\left(1 - \beta + 2\beta \cdot x_i\right)\theta_{\lceil i/2 \rceil}$$

3. $\left((1-\beta)\, r_1, \text{AUX}_1 : g_1\right)$ and $\left((1-\beta)\, r_2, \text{AUX}_1 : g_2\right)$ traders

Let $M_1$ be the total amount that these traders spend on $g_1$. Then

$$M_1 \quad = \sum_{i \in \mathcal{N}_{\overline{G}}(1)} \Delta P_{i,1} \left(1 - \beta + 2\beta \cdot x_i\right) \theta_{\lceil i/2 \rceil} + \left(1 - \beta\right) r_1 \pi \left(\text{AUX}_1\right)$$

Normalizing the prices such that $\frac{1}{n} \sum_{i=1}^{n} \theta_i = 1/\Delta$, this means that

$$M_1 \geq 2\left(1 - \beta\right) + 2\beta \mathbf{x}^\top \cdot \mathbf{P}_1 - O_N\left(\delta\right)$$

Similarly,

$$M_2 \leq 2\left(1 - \beta\right) + 2\beta \mathbf{x}^\top \cdot \mathbf{P}_2 + O_N\left(\delta\right)$$

Therefore,

$$M_1 \geq M_2 + 2\beta\epsilon - O_N\left(\delta\right) = M_2 + \Theta_N\left(\beta\epsilon\right)$$

so the difference between the demands for $g_1$ and $g_2$ from these traders is

$$\frac{M_1}{\pi\left(g_1\right)} - \frac{M_2}{\pi\left(g_2\right)} \geq \frac{M_2 + \Theta_N\left(\beta\epsilon\right)}{\pi\left(g_1\right)} - \frac{M_2\left(1 + \alpha_0\right)}{\pi\left(g_1\right)\left(1 - \alpha_0\right)} = \Theta_N\left(\beta\epsilon\right)$$

Thus the price regulating traders $T_1$ and $T_2$ must have different demands for $g_1$ and $g_2$ - but this can only happen when

$$\frac{1 + \alpha_0}{\pi\left(g_1\right)} = \frac{1 - \alpha_0}{\pi\left(g_2\right)}$$

Therefore, by consecutive applications of Lemma 14,

$$\frac{1 + \beta}{\pi\left(h_1\right)} = \frac{1 - \beta}{\pi\left(h_2\right)}$$

Finally, by (12) this implies that $x_1 = 1$ and $x_2 = 0$.

□